\documentclass[aps,amsmath,prd,floats,floatfix,notitlepage,twocolumn,
superscriptaddress,nofootinbib,showpacs,longbibliography]{revtex4-1}
\usepackage[T1]{fontenc}
\usepackage[utf8]{inputenc}
\usepackage{verbatim}
\usepackage{physics}
\usepackage[dvipsnames, usenames]{xcolor}
\definecolor{linkcolor}{rgb}{0.6,0.0,0.0}
\definecolor{applegreen}{rgb}{0.55, 0.71, 0.0}
\usepackage[hypertexnames=false, unicode, colorlinks=true, linkcolor=linkcolor,
citecolor=linkcolor, filecolor=linkcolor,urlcolor=linkcolor,
pdfusetitle]{hyperref}
\usepackage[all]{hypcap}
\usepackage{graphicx}
\usepackage{xspace}
\usepackage{amssymb}
\usepackage[normalem]{ulem} 
\usepackage{bm} 
\usepackage{orcidlink}
\usepackage{microtype}
\usepackage{lmodern}
\usepackage{blindtext}
\usepackage{cleveref}
\crefname{equation}{Eq.}{Eqs.}
\crefname{section}{Section}{Sections}
\crefname{appendix}{Appendix}{Appendices}
\crefname{figure}{Figure}{Figures}
\crefname{table}{Table}{Tables}
\usepackage{array}
\usepackage[none]{hyphenat}
\usepackage[english]{babel}
\usepackage[title]{appendix}
\usepackage{subcaption}
\usepackage{multirow}
\usepackage{siunitx}
\usepackage{nicematrix}
\graphicspath{%
  {figs/}%
}

\newcommand{\av}[1]{\textcolor{OliveGreen}{{\it\textbf{}}}}

\begin{document}

\newcommand{\ICTS}{\affiliation{International Centre for Theoretical Sciences, Tata Institute of Fundamental Research, Bangalore 560089, India}}
\newcommand{\CITA}{\affiliation{Canadian Institute for Theoretical Astrophysics, University of Toronto, 60 St. George Street, Toronto, ON M5S 3H8, Canada}}

\title{Rapid inference of gravitational-wave signals in the time domain using a heterodyned likelihood}

\author{Neha Sharma~\orcidlink{0009-0005-2928-5411}}
\email{neha.sharma@icts.res.in}
\ICTS{}

\author{Aditya Vijaykumar~\orcidlink{0000-0002-4103-0666}}
\CITA{}

\author{Prayush Kumar~\orcidlink{0000-0001-5523-4603}}
\ICTS{}

\date{\today}

\begin{abstract}
\noindent 
Parameter estimation of gravitational wave signals is computationally intensive and typically requires millions of likelihood evaluations to construct posterior probability distributions. This computational cost increases significantly in the time domain, which requires non-diagonal covariance matrices to compute the likelihood. Consequently, parameter estimation of long-duration gravitational wave signals, such as binary neutron star mergers, becomes computationally infeasible in time domain. In this work, we detail a framework for the heterodyned likelihood that enables rapid inference in the time domain. Our method is applicable to signals with arbitrary mode content, and leverages the smoothness of the ratio of complex-valued waveform modes, approximating the ratio as a linear function within appropriately chosen time bins. This allows downsampling of the waveform modes and a reformulation of the likelihood, such that it depends only on the bin edges. We demonstrate that this likelihood recovers posteriors that are indistinguishable from those obtained using the standard likelihood in the time domain. We also observe dramatic improvement in speed---for a 128 seconds-long gravitational wave signal, our method is at least ${\sim}400$ times faster than the standard time-domain analysis, reducing the wall-clock time to just a few hours. We also demonstrate the reliability and unbiasedness of the likelihood using percentile-percentile tests for binary black hole and binary neutron star injections. We use the Gohberg–Semencul representation of the inverse of Toeplitz covariance matrix to accelerate matrix–vector products, which has potential applications even in non-heterodyned time-domain inference.

\end{abstract}
\maketitle
\section{Introduction}
While a wide variety of astrophysical and cosmological phenomena can produce gravitational waves (GWs)~\cite{LIGOScientific:2018mvr, LIGOScientific:2020ibl, KAGRA:2021vkt,Powell:2024bnp,Abdikamalov:2020jzn,Mezzacappa:2020lsn,Powell:2024bnp,Bonazzola:1995rb,Prix2009,LIGOScientific:2025kei,Wette:2023dom,Pagliaro:2023bvi,Powell:2024bnp,Prasad:2025klz,Caprini:2018mtu,Powell:2024bnp,PhysRevD.97.123505}, current ground based detectors are sensitive to GW signals from coalescing compact binaries~\cite{LIGOScientific:2025slb}. These systems, comprising binary black holes (BBH)~\cite{LIGOScientific:2016aoc}, binary neutron stars (BNS)~\cite{LIGOScientific:2017vwq}, and neutron star-black hole (NSBH) binaries~\cite{LIGOScientific:2021qlt} carry rich information about the source properties, e.g., component masses and spins. These inferred source properties are typically extracted using Bayesian inference~\cite{Thrane:2018qnx,Christensen:2022bxb} and are used in various analyses such as tests of general relativity~\cite{LIGOScientific:2016lio,LIGOScientific:2020tif,LIGOScientific:2021sio,LIGOScientific:2018dkp,LIGOScientific:2019fpa,Krishnendu:2021fga,Perkins:2021mhb,Okounkova:2021xjv,Gupta:2024gun} (GR), understanding the population of compact binaries~\cite{pierro2016binary,LIGOScientific:2016hpm,LIGOScientific:2018jsj,KAGRA:2021duu,LIGOScientific:2020kqk,LIGOScientific:2025pvj}, determination of the equation of state of neutron stars~\cite{LIGOScientific:2017ync,LIGOScientific:2018hze,LIGOScientific:2018cki}, measurement of the Hubble constant~\cite{LIGOScientific:2021aug,LIGOScientific:2025jau,LIGOScientific:2017adf,DES:2019ccw,DES:2020nay} and for triggering multi-messenger astronomical observations~\cite{LIGOScientific:2017ync}.

Parameter estimation of compact binaries is a high-dimensional problem that requires stochastic sampling algorithms~\cite{sharma2017markov,skilling2006nested,sivia2006data} to construct posterior probability distributions. However, these techniques are computationally expensive, often requiring millions of likelihood evaluations. To make these analyses tractable, the majority of parameter estimation anaylses are performed in frequency domain, where stationary noise assumption diagonalises the covariance matrix~\cite{LIGOScientific:2019hgc,Christensen:2022bxb} and simplifies the likelihood computation. In contrast, time-domain analyses require a non-diagonal covariance matrix, making likelihood computation substantially slower (see Section 2 of \cite{Cornish:2020odn} for a detailed discussion). 


While frequency-domain methods are convenient for handling stationary noise, they require windowing the data, which can produce biased posteriors~\cite{Talbot:2021igi,Talbot:2025vth}. Furthermore, the Fourier transform delocalizes the temporal features of source dynamics, making it difficult to isolate a specific part of the GW signal. Time-domain inference, however, does not require windowing data and can be used to analyze specific parts of a signal. This capability makes it immensely useful in various applications such as the tests of GR, specifically the inspiral-merger-ringdown consistency test (IMRCT)~\cite{Ghosh:2016qgn,Ghosh:2017gfp,Breschi:2019wki,Shaikh:2024wyn,Krishnendu:2021fga}, studies of BBH ringdown~\cite{Isi:2021iql,Siegel:2024jqd} and the area theorem~\cite{Isi:2020tac,LIGOScientific:2025rid}\footnote{See also Ref.~\cite{Kou:2025bhk} for a recent comparison of time- and frequency-domain parameter estimation with applications to searches for the stochastic background.}. These analyses require a clear separation of the inspiral and merger-ringdown parts of the signal. While the time of merger is clearly measured and can act as a separator in time domain, identifying a cut-off frequency that separates the pre-merger and post-merger parts of the signal in the frequency domain is generally not possible, especially in the presence of significant higher harmonics~\cite{Mills:2020thr,Barausse:2011kb,Garcia-Quiros:2020qpx,Kankani:2025dvz}. 
A similar complication arises on the inclusion of orbital eccentricity. In eccentric binaries, the frequency doesn't increase monotonically with time. Instead, it shows an oscillatory behaviour around orbit-averaged frequency~\cite{Shaikh:2023ypz}. This makes it challenging to determine a clear cut-off frequency as higher frequencies may overlap with the inspiral part and lower frequencies may overlap with the merger and ringdown part of the signal \cite{Sinha:2025vmc}.  While biases in tests of GR could be negligible for binaries detected in current ground based detectors such as Advanced LIGO~\cite{LIGOScientific:2014pky}, Advanced Virgo~\cite{VIRGO:2014yos} and KAGRA~\cite{KAGRA:2020tym,Somiya:2011np,Aso:2013eba}, where most of the eccentricity is radiated away before GW signal enters the detector's sensitive band, it can be significant for the binary systems that remain eccentric close to the merger. This problem can be solved by using a well-defined time of merger as a separator in the time domain. 

Beyond separating various phases of a GW signal, time domain inference also enables us to identify the origin of information about various parameters in the strain data~\cite{Miller:2023ncs,Udall:2024ovp,Miller:2025eak}. This method has been used to trace the contribution of each cycle to precession constraints~\cite{Miller:2023ncs, Miller:2025eak}, as well as to study the effects of overlapping glitches on parameter measurements~\cite{Udall:2024ovp}.

Despite these advantages, parameter estimation in time-domain is computationally expensive. In time domain, a single likelihood evaluation can take up to a few seconds for a network of Advanced LIGO~\cite{LIGOScientific:2014pky}, Advanced Virgo~\cite{VIRGO:2014yos} and KAGRA~\cite{KAGRA:2020tym,Somiya:2011np,Aso:2013eba} detectors, and a typical parameter estimation run requires $\mathcal{O}(10^7)$ such evaluations. This computational cost is expected to increase for next-generation GW detectors~\cite{dj7k-tk37} such as Einstein Telescope~\cite{Punturo:2010zz} and Cosmic Explorer~\cite{LIGOScientific:2016wof}, which will detect longer GW signals. While numerous methods have been developed to accelerate the parameter estimation in frequency domain~\cite{canizares2015accelerated,smith2016fast,morisaki2020rapid,morisaki2021accelerating,morisaki2023rapid,vinciguerra2017accelerating,pathak2023fast,pathak2024prompt,edwards2023ripple,wong2023fast,cornish2010fast,cornish2021heterodyned,zackay2018relative,Krishna:2023bug,cornish2010fast,cornish2021heterodyned,zackay2018relative,Krishna:2023bug,finstad2020fast,narola2023relative,leslie2021mode,Dai:2018dca}, similar efforts in time domain have been lacking.

In this work, we present a framework for fast likelihood evaluation in time domain, which can reduce the time required for a full parameter estimation run by at least an order of magnitude. Our method employs the principles of \textit{heterodyning}, also known as \textit{relative binning}~\cite{cornish2010fast,cornish2021heterodyned,zackay2018relative,Krishna:2023bug,finstad2020fast,narola2023relative,leslie2021mode,Dai:2018dca}. The fundamental idea behind this method is that gravitational waveforms from compact binaries are smooth functions of their source parameters~\cite{cornish2010fast,Field:2013cfa,zackay2018relative}. Therefore, the waveforms generated at parameters confined to a local region in parameter space are fairly similar, and the ratio of a sampled waveform and a pre-computed fiducial waveform is a smooth, slowly varying function of frequency. This ratio can be approximated as a linear curve in small frequency bins. Since any line is fully determined by its endpoints, the waveforms only need to be generated at the bin edges. This significantly downsamples the waveform, resulting in faster waveform generation. The full waveform, if needed, can be reconstructed by interpolating this linear curve and multiplying it by the fiducial waveform.  The likelihood is also expressed in terms of the ratio of waveforms at bin edges and precomputed terms within each bin, which depend on the fiducial waveform, noise properties, and the data. As a result, the likelihood calculation only requires calculating the waveform on the bin edges, making the calculation much faster.

We use a similar approach in time domain, with a key difference: the heterodyning approximation is applied to the complex-valued waveform multipoles $h^{\ell m}(t)$ instead of the individual polarizations $h_{+, \times}(t)$. In the frequency domain, the individual polarizations as well as the multipoles are complex-valued, so their ratio is well defined at every point. In contrast, the ratio of real-valued polarizations in the time domain can be undefined at zero-crossings, but the ratio of complex-valued waveform modes is well-behaved across the parameter space and at each time point. We show that the ratio of sampled waveform modes and fiducial waveform modes can be approximated as a (complex) linear curve in time bins, effectively downsampling the waveform. The likelihood then depends only on this ratio evaluated at bin edges, and pre-computed terms (known as \textit{summary data}) that depend on the fiducial modes, detector data and noise covariance matrix. The speedup due to this approximation stems from (i) downsampling the waveform, and (ii) avoiding large matrix products at each likelihood calculation. This is especially effective for longer signals because the computational cost becomes decoupled from the signal duration—the number of bins stays nearly constant even as the total data points grow linearly.

The efficiency and accuracy of our method relies on the choice of bins. The bins are wider during the early inspiral part, with their size gradually reducing close to the merger. These bin sizes are calculated using a binning criterion derived in \cref{s2:Binning_criteria}. For merger and early ringdown, the bins are chosen to be small and uniform in size over a small time interval near the merger. For the late ringdown part, the bins are uniform but wider. 

We validate our framework through an injection study in simulated Gaussian noise.
For a 2-second long GW signal, time-domain heterodyning can accurately recover the posterior probability distributions that are indistinguishable from those obtained using the full likelihood method\footnote{We refer to the usual likelihood used in standard time-domain  analyses and defined in \cref{eq:TD_likelihood,eq:TD_likelihood_single_detector}, as ``full likelihood'' throughout this paper.}.
Heterodyning evaluates the likelihoods of a 2-second-long signal $\sim$19 times faster than the full likelihood method. This speedup scales dramatically with signal duration. For signals with durations of 16 and 128 seconds, a single likelihood evaluation using heterodyning takes only 6.4 and 9.2 ms, respectively. This corresponds to speedup factors of 124 and 340 compared to the full likelihood. Thus, analyses using the heterodyned likelihood can be completed in a few hours, whereas equivalent analyses with the full likelihood would require weeks to months. In all cases, the recovered posteriors are consistent with the injection parameters. 
To further test our method, we make percentile-percentile (p-p) plots for a simulated population of BBH and BNS systems and recover unbiased posteriors. 

This paper is organised as follows. In \cref{s1:framework}, we describe the conceptual framework of heterodyning in the time domain. We first outline Bayesian parameter estimation in \cref{s2:bayesian_pe}, then introduce heterodyning approximation in \cref{s2:Relative_binning}, followed by the derivation of summary data in \cref{s2:Summary_data} and binning criteria in \cref{s2:Binning_criteria}. In \cref{s1:Results}, we examine the accuracy of our framework by comparing it with standard Bayesian analysis and by presenting p-p plots for signals of various durations. We also discuss the speedup factors achieved in each analysis.

\section{Conceptual framework}\label{s1:framework}
\subsection{Time-domain likelihood}\label{s2:bayesian_pe}
Bayesian inference allows us to estimate physical parameters of the GW signals recorded in our detectors~\cite{Thrane:2018qnx}. With this tool, we can localize the sources in the sky, determine their component masses, spins and orientation angles. These inferred properties are then used in a wide variety of downstream analyses to understand astrophysical and cosmological aspects of these signals. 

The goal of Bayesian inference is to estimate the posterior probability distribution $p(\vec{\theta}|d,M)$ of source parameters $\vec{\theta}$, given detector strain $d$ and a signal model $M$. According to Bayes' theorem, this posterior probability distribution is given by
\begin{equation}
    \begin{split}
        p(\vec{\theta}|d,M) &= \frac{\mathcal{L}(d|\vec{\theta},M)\pi(\vec{\theta}|M)}{\int\mathcal{L}(d|\vec{\theta},M)\pi(\vec{\theta}|M) d\vec{\theta}}\\ &\equiv \frac{\mathcal{L}(d|\vec{\theta},M)\pi(\vec{\theta}|M)}{\mathcal{Z}(M)},
    \end{split}
\end{equation}
where $\pi(\vec{\theta}|M)$ is the prior probability distribution of parameters $\vec{\theta}$, which is chosen based on known physical constraints on the source parameters. The term $\mathcal{L}(d|\vec{\theta},M)$ is the likelihood of data $d$, given parameters $\vec{\theta}$. The denominator, $\mathcal{Z}(M)$, is the \textit{Bayesian evidence}, which acts as a normalization factor. The Bayesian evidence indicates whether the observed data is well described by model $M$, and can be used to compare competing models. 

In a network of $K$ detectors, the data from $k^{\mathrm{th}}$ detector is modelled as
\begin{equation}
    d^{(k)}(t) = s^{(k)}(t;\vec{\theta},M) + n^{(k)}(t),
\end{equation}
where $s^{(k)}(t;\vec{\theta},M)$ is the gravitational waveform for parameters $\vec{\theta}$ predicted by the model $M$, and $n^{(k)}(t)$ is the noise time series for $k^{\mathrm{th}}$ detector. Assuming the noise in each detector is independent, the total log-likelihood is the sum of log-likelihood for each detector:
\begin{equation}
    \begin{split}
        \text{ln}\mathcal{L}(d|\vec{\theta},M) &= \sum_{k=1}^K\text{ln}\mathcal{L}(d^{(k)}|\vec{\theta},M).
    \end{split}\label{eq:TD_likelihood}
\end{equation}
For Gaussian noise, the log-likelihood for $k^{\mathrm{th}}$ detector is
\begin{equation}
    \begin{split}
        \text{ln}\mathcal{L}(d^{(k)}|\vec{\theta},M) = &-\frac{1}{2}\sum_{i,j = 0}^{N-1}(d^{(k)}_i-s^{(k)}_i)(C^{(k)})^{-1}_{ij}(d^{(k)}_j-s^{(k)}_j) \\ 
        &+ \text{constant},
    \end{split}
    \label{eq:TD_likelihood_single_detector}
\end{equation}
where $C^{(k)}$ is covariance matrix for $k^{\mathrm{th}}$ detector. For stationary Gaussian noise, the covariance matrix has a symmetric Toeplitz structure:
\begin{equation}
    C_{ij} = \rho(|i-j|),
\end{equation}
where $\rho$ is the autocorrelation function (ACF). The ACF can be estimated by autocorrelating a long segment of noise or by taking the inverse Fourier transform of the noise power spectral density~\cite{Isi:2021iql,Siegel:2024jqd,Wang:2024liy}. 

The waveform $s^{(k)}(t;\vec{\theta})$ for a set of physical parameters $\vec{\theta}$ is the projection of two GW polarizations $h_+, h_\times$ onto the detector:
\begin{equation}
\begin{split}
    s^{(k)}(t;\vec{\theta}) = \;&F_+^{(k)}(t;\alpha,\delta,\psi)h_+(t;\vec{\theta})
    \\ +&F_\times^{(k)}(t;\alpha,\delta,\psi) h_\times(t;\vec{\theta}),
 \end{split}\label{eq:s_theta}
\end{equation}
where $F_+^{(k)}, F_\times^{(k)}$ are antenna pattern functions of the $k^{\mathrm{th}}$ detector, which depend on the source's sky location (right ascension $(\alpha)$ and declination ($\delta$) angle), and the polarization angle ($\psi$)~\cite{Isi:2022mbx}, which is the third Euler angle required to rotate the source frame to the detector-aligned frame. The antenna pattern functions also depend on the detector's location and orientation, and they measure the detector's sensitivity to different directions and polarization of the GW signal. For short-duration transients, the Earth's rotation can be neglected, and these functions can be treated as constant in time.

The GW polarizations $h_+, h_\times$ can be decomposed into waveform modes $h^{\ell m}$, and spin-weighted spherical harmonics $_{-2}Y^{\ell m}$ as 
\begin{equation}
    \begin{split}
        h_+ - i h_\times = \sum_{\ell = 2}^{\infty}\sum_{m = -\ell}^{\ell} h^{\ell m}(t;\boldsymbol{\lambda}, D_L)  \;_{-2}Y^{\ell m}(\iota,\phi) \equiv \mathcal{H}.
    \end{split}\label{eq:pol_in_terms_of_modes}
\end{equation}
The spin-weighted spherical harmonics $_{-2}Y^{\ell m}$ depend on the inclination angle, $\iota$, between the binary's orbital angular momentum and the line of sight, and the orbital phase at coalescence $\phi_c$. The waveform modes $h^{\ell m}$ depend on the luminosity distance, $D_L$ and intrinsic parameters $\boldsymbol{\lambda}\equiv\{m_1,m_2,\vec{S}_1, \vec{S}_2\}$. Note that the source's sky location ($\alpha, \delta$) and polarization angle ($\psi$) do not affect the waveform modes and spin-weighted spherical harmonics. They determine the detector response via antenna pattern functions [\cref{eq:s_theta}].

\begin{figure*}[htb]
     \begin{subfigure}[b]{0.47\textwidth}
         \centering
         \includegraphics[width=\linewidth]{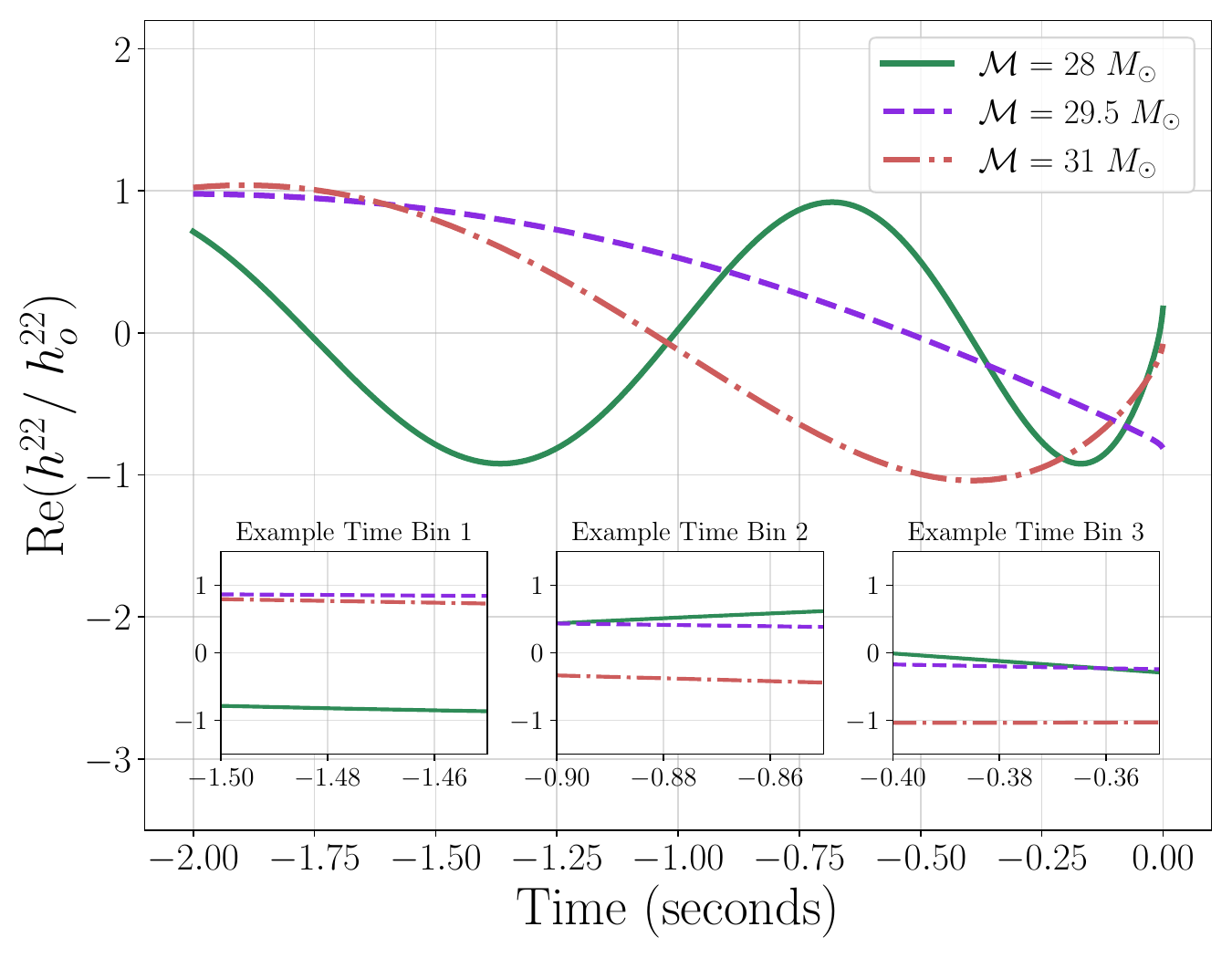}
         \caption{}
         \label{fig:22_mode_ratio}
     \end{subfigure}
     \hfill
     \begin{subfigure}[b]{0.47\textwidth}
         \centering
         \includegraphics[width=\linewidth]{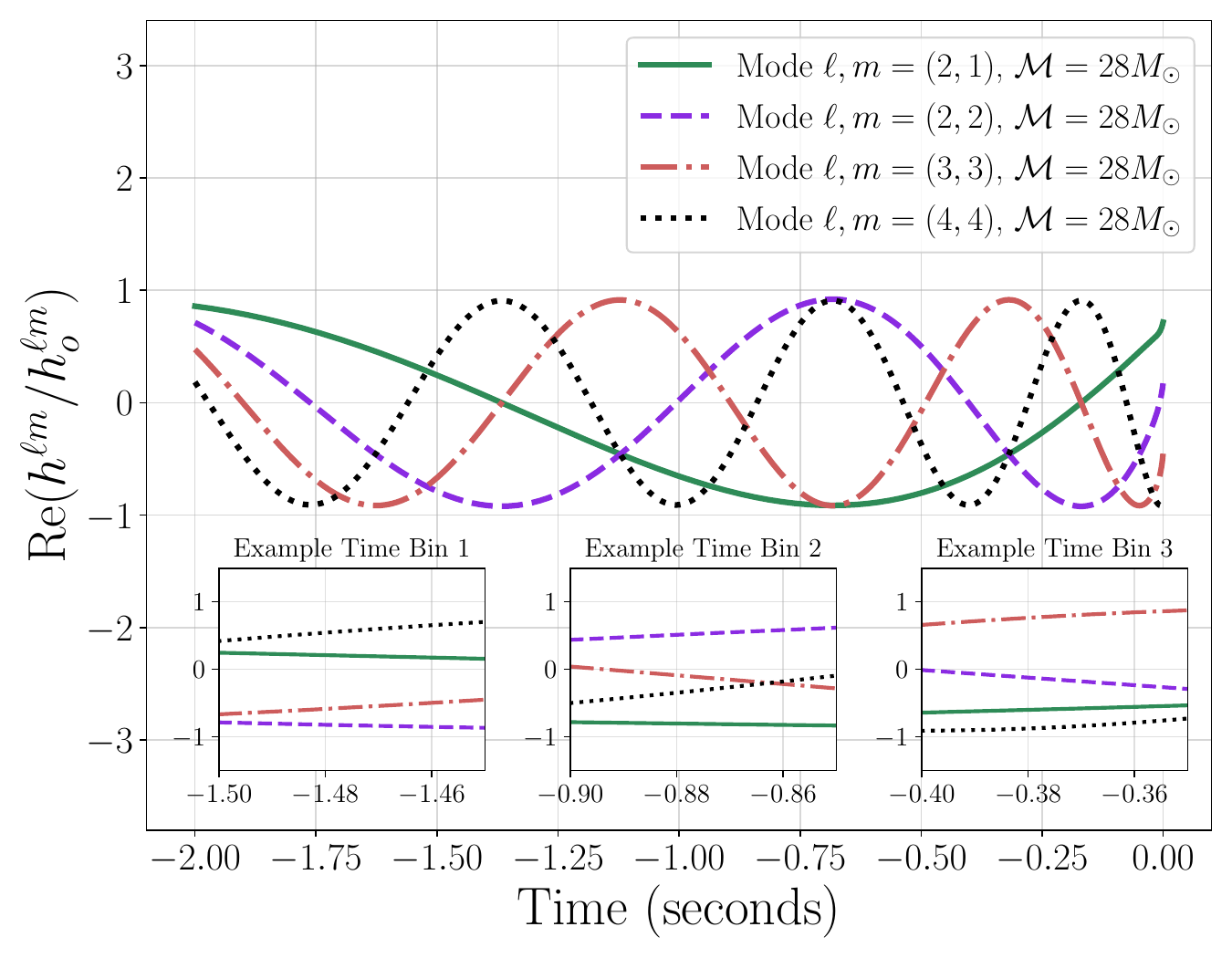}
         \caption{}
         \label{fig:lm_mode_ratio}
     \end{subfigure}
     \caption{The ratio of sampled to fiducial (reference) waveform modes, $r^{\ell m} = h^{\ell m}/h^{\ell m}_o$, behaves as a slowly varying function of time. (a) The ratio for (2,2) mode, $r^{22}(t)$, plotted for different sampled chirp masses. (b) The ratio for various $(\ell, m)$ modes, $r^{\ell m}(t)$, at a fixed sampled chirp mass of $28\;M_\odot$. The insets in both figures show three small time bins, demonstrating that the ratio $r^{\ell m}(t)$ can be approximated as a straight line within each time bin. In both panels, the fiducial chirp mass is fixed at $30 \; M_\odot$.
     }\label{fig:ratios_of_waveform.}
\end{figure*}

For a quasi-circular system, the full set of parameters $\vec{\theta}$ defining a GW signal can be divided into intrinsic and extrinsic parameters. The intrinsic parameters include masses ($m_1,m_2$), spins ($\vec{S}_1,\vec{S}_2$), and the extrinsic parameters include luminosity distance ($D_L$), coalescence time ($t_c$), inclination angle ($\iota$), phase at coalescence ($\phi_c$), right ascension ($\alpha$), declination ($\delta$) angle, and the polarization angle ($\psi$). In this work, we sample all 11 parameters of aligned spin systems. Instead of sampling component masses and spin vectors, we sample chirp mass,
\begin{equation}
    \mathcal{M} = \frac{(m_1 m_2)^{3/5}}{(m_1+m_2)^{1/5}},
\end{equation} 
mass ratio, 
\begin{equation}
    q = \frac{m_2}{m_1},
\end{equation}
 and aligned spins ($\chi_1, \chi_2$). In extrinsic parameters, we sample the arrival time at H1 detector ($t_\mathrm{H1}$) instead of the geocentric time.

To simplify the notation, we hereafter omit the detector index $k$ in the likelihood expression. 

\subsection{Heterodyning in Time Domain}\label{s2:Relative_binning}

The fundamental idea behind heterodyning is that the waveform modes $h^{\ell m}(\vec{\theta})$ vary smoothly with the model parameters $\vec{\theta}$ ~\cite{cornish2010fast,cornish2021heterodyned,zackay2018relative,Krishna:2023bug}. Therefore, $h^{\ell m}(\vec{\theta})$ can be described as a small perturbation of a waveform modes calculated at a nearby reference set of parameters $\vec{\theta}_o$, also known as \textit{fiducial parameters}. The corresponding waveform modes are referred to as \textit{fiducial waveform modes}, $h^{\ell m}_o$. 

The ratio of sampled waveform modes and fiducial waveform modes is a slowly varying function of time. Within a small time bin,  this function can be well approximated by a linear curve, as illustrated in \cref{fig:22_mode_ratio,fig:lm_mode_ratio}. Within a specific bin, $b$, this ratio can be expressed as  
    \begin{equation}
        \begin{split}
            r^{\ell m}(b)  =\; \frac{h^{\ell m}}{h_o^{\ell m} } &=r_o^{\ell m}(b)+r_1^{\ell m}(b)(t-t_m(b)),\\
        \end{split}\label{eq:ratio_of_waveforms}
    \end{equation}
where     
    \begin{equation}
        \begin{split}
            r^{\ell m}_o(b)& = \frac{r^{\ell m}\left(t_f(b)\right) + r^{\ell m}\left(t_i(b)\right)}{2},\\
            r^{\ell m}_1(b)& = \frac{r^{\ell m}\left(t_f(b)\right) - r^{\ell m}\left(t_i(b)\right)}{t_f(b) - t_i(b)}.
        \end{split}\label{eq:Ratio_coeff}
    \end{equation}
Here, $t_m(b)$ is the \textit{centre} of each time bin $b$, and $t_i(b)$ and $t_f(b)$ are initial and final times of the bin, i.e. the bin edges. The coefficients $r_o^{\ell m}(b)$ and $r_1^{\ell m}(b)$ are constant within each bin and are computed separately for each $(\ell ,m)$ mode.

This formalism provides a significant computational advantage. As \cref{eq:Ratio_coeff} suggests, the coefficients $r_o^{\ell m}$ and $r_1^{\ell m}$ need to be computed only at the bin edges. This means that the sampled waveform modes $h^{\ell m}$ also need to be generated only at the bin edges, instead of the full time grid. The waveform modes $h^{\ell m}$ at full time grid, if needed, can be computed using the relation,
\begin{equation}
    \begin{split}
        h^{\ell m} = h^{\ell m}_o(r_o^{\ell m}(b)+r_1^{\ell m}(b)(t-t_m(b))).
    \end{split}\label{eq:hlm_using_ratio}
\end{equation}
This approach helps in downsampling of waveform modes, making waveform generation significantly faster. For example, for $150$ time bins, any waveform modes need to be generated at only $150$ time points instead of $T\times f_s$ points, where $T$ is the duration of GW signal in seconds and $f_s$ is the sampling frequency. This becomes increasingly efficient for longer GW signals, where generating full waveform modes, i.e. at full time grid, can take from a few hundred milliseconds to several seconds. With heterodyning, waveform generation time reduces to a few milliseconds, even for longer GW signals.

The downsampling of waveform modes is the first step toward accelerating parameter estimation. The second step involves making the likelihood computation more efficient. This is achieved by rewriting the likelihood defined in \cref{eq:TD_likelihood_single_detector} in terms of the ratio of modes $r^{\ell m}$ (see \cref{eq:ratio_of_waveforms}) evaluated only at bin edges, and a set of pre-computed quantities we refer to as \textit{summary data}. As we will show in the next subsection, the \textit{summary data} depends only on the fiducial waveform modes $h^{\ell m}_o$, the covariance matrix, $C$, and data $d$. These steps substantially simplify the likelihood computation: (i) likelihood evaluation during sampling requires no matrix multiplications, and (ii) the evaluation time scales with the number of bins instead of the total number of time samples, yielding a dramatic speedup.  We continue this discussion in \cref{s2:Summary_data}, where we also provide explicit expressions for \textit{summary data}, and heterodyned likelihood.

\subsection{Summary data}\label{s2:Summary_data}
In this section, we reformulate the likelihood defined in \cref{eq:TD_likelihood_single_detector} in terms of \textit{summary data} and ratio of waveform modes, $r^{\ell m}$. We first derive the analytic form of \textit{summary data} for waveforms containing only dominant waveform modes. This derivation is then extended to multimodal waveforms in \cref{apx:Summary_data_subdominant_modes}.

The likelihood defined in \cref{eq:TD_likelihood_single_detector} can be decomposed as
\begin{equation}
\begin{split}
     \text{ln} \mathcal{L}(d|\vec{\theta}) =
     &-\frac{1}{2}\underbrace{\sum_{i,j = 0}^{N-1}d_iC_{ij}^{-1}d_j}_{L_1(d,d)}+\underbrace{\sum_{i,j = 0}^{N-1}d_iC_{ij}^{-1}s_j}_{L_1(d,s)}\\&-\frac{1}{2}\underbrace{\sum_{i,j = 0}^{N-1}s_iC_{ij}^{-1}s_j}_{L_1(s,s)}.
\end{split} \label{eq:decomposed_likelihood}
\end{equation}
The first term, $L_1(d,d)$, depends only on the data and the covariance matrix,  and is therefore computed only once. The second and third terms, $L_1(d,s)$ and $L_1(s,s)$ respectively, depend on the waveform, $s$, and must be evaluated repeatedly during sampling. To make these terms computationally efficient, we express the waveform, $s$, in terms of the fiducial waveform modes $h^{\ell m}_o$ and the ratio $r^{\ell m}$. Using the relations defined in \cref{eq:s_theta,eq:pol_in_terms_of_modes,eq:ratio_of_waveforms,eq:Ratio_coeff,eq:hlm_using_ratio}, the term $L_1(d,s)$ for dominant waveform modes, $(l,m) = (2,\pm 2)$, can be written as
\begin{equation}
    \begin{split}
        L_1(d,s)= F_+ \mathfrak{Re}\big[& _{-2}Y^{22} L_1(d,h^{22}) \\ & + \;_{-2}Y^{2,-2} L_1(d,h^{2,-2})\big]\\ - F_\times \mathfrak{Im}\big[&_{-2}Y^{22} L_1(d,h^{22})\\& + \;_{-2}Y^{2,-2} L_1(d,h^{2,-2})\big].
    \end{split}\label{eq:L_1_d_s_22_1}
\end{equation}
By using the relation $h^{2,-2} = (h^{22})^*$, this expression simplifies to
\begin{equation}
    \begin{split}
        L_1(d,s)= F_+ \mathfrak{Re}\big[&_{-2}Y^{22} L_1(d,h^{22})\\ & + \;_{-2}Y^{2,-2} L_1(d,h^{22})^*\big]\\ - F_\times \mathfrak{Im}\big[&_{-2}Y^{22} L_1(d,h^{22})\\ & + \;_{-2}Y^{2,-2} L_1(d,h^{22})^*\big],
    \end{split}\label{eq:L_1_d_s_22_2}
\end{equation}
where 
\begin{equation}
    \begin{split}
        L_1(d,h^{22})& = \sum_b [r_o^{22}(b) A_o^{22}(b) + r_1^{22}(b) A_1^{22}(b)].
    \end{split}\label{eq:L_d_h22}
\end{equation}
Similarly, the term $L_1(s,s)$ can be expressed as  
\begin{equation}
    \begin{split}
        L_1(s,s) = &\;\frac{F_+^2+F_\times^2}{2}\;\mathfrak{Re}( L_1(\mathcal{H}^*,\mathcal{H}))
        \\&+\frac{F_+^2-F_\times^2}{2}\;\mathfrak{Re}(L_1(\mathcal{H},\mathcal{H}))\\
        &\; -F_+F_\times \mathfrak{Im}\left[L_1(\mathcal{H},\mathcal{H})+L_1(\mathcal{H}^*,\mathcal{H})\right],
    \end{split}\label{eq:L_1_s_s22}
\end{equation}
where
\begin{equation}
    \begin{split}
        L_1(\mathcal{H},\mathcal{H}) =\; &_{-2}Y^{2,2} \;_{-2}Y^{2,2}L_1(h^{22},h^{22})  \\  &+_{-2}Y^{2,-2} \;_{-2}Y^{2,-2}L_1(h^{22},h^{22})^* 
        \\& +2( _{-2}Y^{2,2} \;_{-2}Y^{2,-2} L_1(h^{22},h^{*\;22})),
    \end{split}\label{eq:L_1_H_H22}
\end{equation}
and 
\begin{equation}
    \begin{split}
        L_1(\mathcal{H}^*,\mathcal{H}) = \; &  _{-2}Y^{*\;2,2} \;_{-2}Y^{2,2} L_1(h^{22},h^{22})^* \\ &+_{-2}Y^{2,-2} \;_{-2}Y^{2,2}L_1(h^{22},h^{22})\\&+(_{-2}Y^{*\;2,-2} \;_{-2}Y^{2,-2}\\&+_{-2}Y^{*\;2,2} \;_{-2}Y^{2,2})L_1(h^{22},h^{*\;22}).
    \end{split}\label{eq:L_1_H*_H22}
\end{equation}
The inner products appearing in these equations are
\begin{equation}
    \begin{split}
        L_1(h^{22},h^{22}) = \sum_{b_1} \sum_{b_2} &[r_o^{22}(b_1) r_o^{22}(b_2) B_o^{22}(b_1,b_2) \\& + (r_o^{22}(b_1)r_1^{22}(b_2) \\&+ r_1^{22}(b_1)r_o^{22}(b_2))B_1^{22}(b_1,b_2)],
\end{split}\label{eq:L_1_h22_h22}
\end{equation}
and
\begin{equation}
\begin{split}
        L_1(h^{*\;22},h^{22}) = \sum_{b_1}\sum_{b_2}&[r_o^{*22}(b_1)r_o^{22}(b_2) B_2^{22}(b_1,b_2) \\&+ (r_o^{*22}(b_1)r_1^{22}(b_2)\\&+r^{*22}_1(b_1)r_o^{22}(b_2))B_3^{22}(b_1,b_2)].
    \end{split}\label{eq:L_1_h*22_h22}
\end{equation}
The computationally expensive terms $A_n^{22}(b)$, and $B_n^{22}(b_1,b_2)$ are pre-computed and are collectively known as \textit{summary data}. These terms are defined as
\begin{equation}
    \begin{split}
        A_o^{22}(b)& = \sum_{j\in b}\sum_{i = 0}^N d_i C_{ij}^{-1} h^{22}_{o,j},\\
        A_1^{22}(b)& = \sum_{j\in b}\sum_{i = 0}^N d_i C_{ij}^{-1} h^{22}_{o,j} (t_j-t_c),\\
        B_o^{22}(b_1,b_2)& = \sum_{i\in b_1}\sum_{j \in b_2} h^{22}_{o,i} C^{-1}_{ij}h^{22}_{o,j},\\
        B_1^{22}(b_1,b_2)& = \sum_{i\in b_1}\sum_{j \in b_2} h^{22}_{o,i} C^{-1}_{ij}h^{22}_{o,j} (t_j-t_c),\\
        B_2^{22}(b_1,b_2)& = \sum_{i\in b_1}\sum_{j \in b_2} h^{*\;22}_{o,i} C^{-1}_{ij}h^{22}_{o,j},\\
        B_3^{22}(b_1,b_2)& = \sum_{i\in b_1}\sum_{j \in b_2} h^{*\;22}_{o,i} C^{-1}_{ij}h^{22}_{o,j}(t_j-t_c).
    \end{split}\label{eq:Summary data_22}
\end{equation}
In these relations, $h^{22}_{o,i}$ is the dominant mode of fiducial waveform evaluated at time $t_i$, i.e. $h^{22}_{o,i}\equiv h^{22}_0(t_i)$ and $N = T\times f_s$, as defined earlier, is the total number of data points. By pre-computing the \textit{summary data} for given time bins, the likelihood can be efficiently computed using \cref{eq:decomposed_likelihood,eq:L_1_d_s_22_1,eq:L_1_d_s_22_2,eq:L_d_h22,eq:L_1_s_s22,eq:L_1_H_H22,eq:L_1_H*_H22,eq:L_1_h22_h22,eq:L_1_h*22_h22}.

We can generalise this framework for subdominant modes and precessing waveforms. In the presence of subdominant modes, we compute summary data (see \cref{apx:Summary_data_subdominant_modes}) and ratio of modes $r^{\ell m}$ individually for each $(\ell, m)$ mode to construct mode-by-mode heterodyned likelihood. For precessing waveforms, we can use the same framework developed in this section, with a larger number of bins to resolve the modulations in the amplitude of the waveform. Alternatively, we can use heterodyning approximation in co-precessing frame and apply frame rotation in subsequent steps.

\subsection{Binning Criteria}\label{s2:Binning_criteria}
\begin{figure*}[htb]
    \centering
    \includegraphics[width=1.0\linewidth]{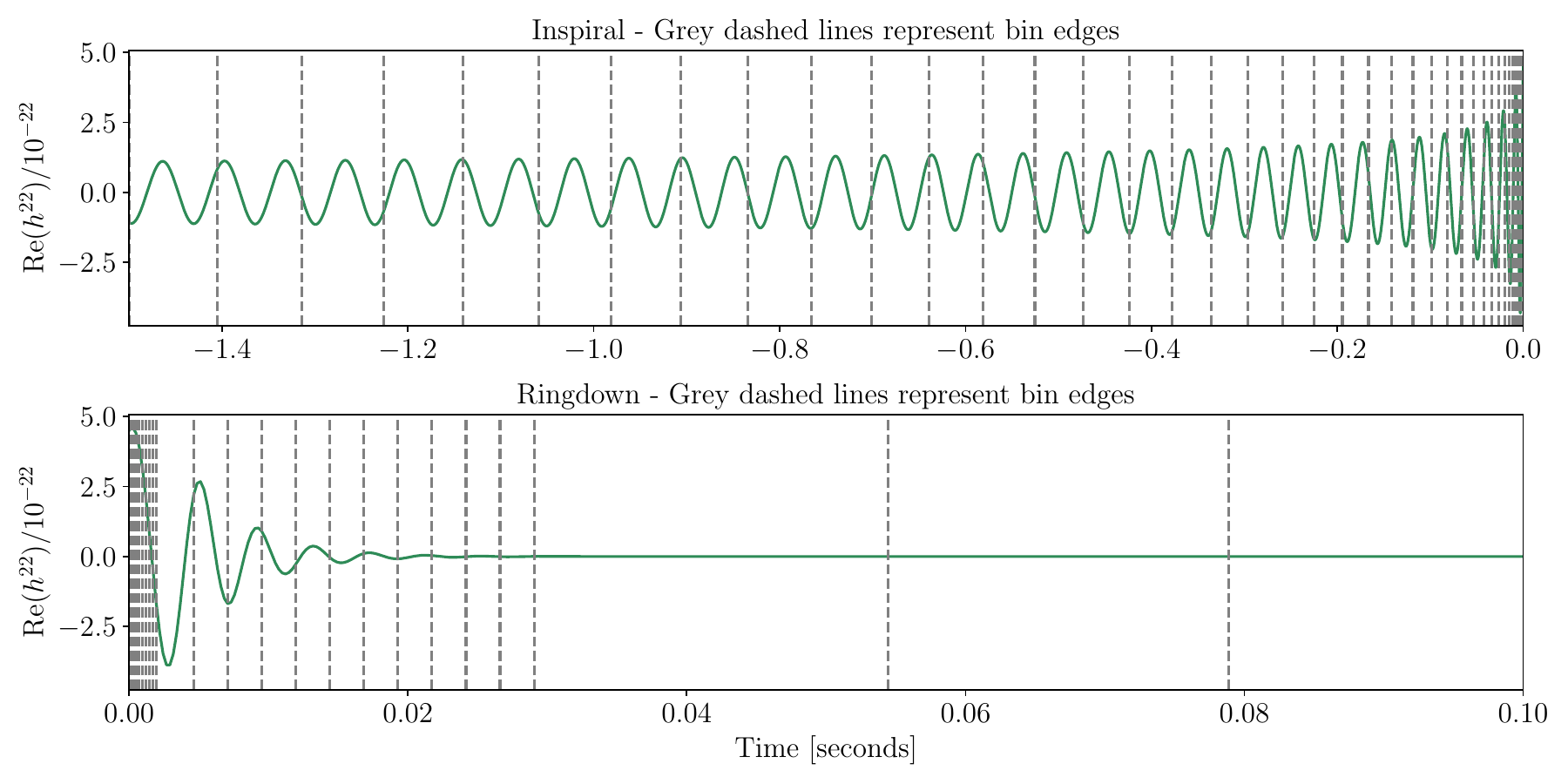} 
    \caption{An illustration of time bins constructed for the inspiral and merger-ringdown part of a GW signal. The upper panel shows that the time bins for early inspiral part are wider. The bin width progressively reduces towards merger, with a few bins very close to merger containing only one data point. The lower panel shows the time bins for merger and ringdown part of signal. Near merger, the bins are narrow, some containing only one data point. The bins widen as the signal decays rapidly during the late ringdown phase.}
    \label{fig:Illustration of time bins}
\end{figure*}
In the heterodyning method, the likelihood is approximated using the \textit{summary data} and the ratio $r^{\ell m}$, both of which are functions of time bins. Thus, the error in likelihood evaluation depends on the choice of bins. The bins must be narrow enough to ensure that the ratio $r^{\ell m}$ is well approximated as a linear function within each bin, but wide enough to keep the total number of bins small.
In this section, we derive a binning criterion which is used to optimise the selection of bins in the time domain. Our approach is analogous to the method described in ~\cite{zackay2018relative,Krishna:2023bug}, which was originally developed for the frequency domain.

We start by defining the polarizations of the fiducial waveform ($h_{o,+}, h_{o,\times}$) and a sampled waveform ($h_+,h_\times$), with parameters close to the fiducial parameters, as
\begin{equation}
\begin{split}
    h_{o,+}(t)-ih_{o,\times}(t) = &\;|\mathcal{H}_o(t)|e^{i\Psi_o (t)},\\ h_+(t) -ih_\times(t) =&\; |\mathcal{H}(t)|e^{i\Psi_o(t) +i\delta\Psi (t)},
\end{split}
\end{equation}
where $\mathcal{H}$ is defined in \cref{eq:pol_in_terms_of_modes}. The polarizations of the sampled waveform differ from the polarizations of the fiducial waveform by a small perturbation in phase, $\delta \Psi(t)$. Since the amplitude $|\mathcal{H}(t)|$ varies slowly compared to the phase, we can assume that for a small perturbation $\delta \Psi (t)$, the amplitudes are approximately equal, i.e. $|\mathcal{H}(t)|\simeq|\mathcal{H}_o(t)|$. For a sufficiently small phase difference, the exponential term can be approximated as $e^{i\delta \Psi(t)} \simeq (1+i\delta \Psi(t))$. This allows us to write the polarizations of the sampled waveform as 
\begin{equation}
    \begin{split}
        h_+(t) =&\; h_{o,+}(t) +h_{o,\times}(t) \delta\Psi(t), \\ h_\times(t)  =&\;  h_{o,\times}(t) -h_{o,+}(t) \delta\Psi(t).
    \end{split}\label{eq:Linearised_pol}
\end{equation}
The corresponding fiducial and sampled waveforms are given by
\begin{equation}
    \begin{split}
        s_{o,i} =&\; F_+h_{o,+,i}+F_\times h_{o,\times,i}\\
        s_{i} =&\; F_+h_{+,i}+F_\times h_{\times,i}
    \end{split}\label{eq:fiducial_sampled_strain}
\end{equation}
where index $i$ denotes the evaluation of various quantities at time $t_i$. By substituting the linearized polarizations defined in \cref{eq:Linearised_pol} into \cref{eq:fiducial_sampled_strain}, the sampled waveform can be written as 
\begin{equation}
    \begin{split}
        s_{i}=&\; s_{o,i}+(F_+h_{o,\times,i}-F_\times h_{o,+,i} )\delta\Psi_i.
    \end{split}
\end{equation}
This shows that the sampled waveform is simply the sum of the fiducial waveform and a small perturbation proportional to $\delta \Psi_i$.

We now examine the effect of this perturbation on the likelihood. Let ln$\mathcal{L}_o$ be the log-likelihood of the fiducial waveform, and ln$\mathcal{L}$ be that of the sampled waveform. The difference between these two log-likelihoods is 
\begin{equation}
    \begin{split}
         \Rightarrow \text{ln}\mathcal{L}_o-\text{ln}\mathcal{L} =\sum_{i,j = 0}^{N-1}&\Big[d_iC_{ij}^{-1}(-F_+h_{o,\times,j}+ F_\times h_{o,+,j})\\&+F_+ s_{o,i}C_{ij}^{-1} h_{o,\times,j}\\&-F_\times s_{o,i}C_{ij}^{-1} h_{o,+,j}\Big] \delta\Psi_j.
    \end{split}\label{diff_likelihood}
\end{equation}
This expression is obtained by neglecting the quadratic terms in $\delta \Psi_i$. The phase $\Psi$ can also be written as a sum of powers of time,
\begin{equation}
    \begin{split}
         \Psi(t_j) =&\; \sum_{k} \alpha_k t_j^{\gamma_k},\\
        \Rightarrow \delta \Psi(t_j) = &\; \sum_{k} (\delta\alpha_k) t_j^{\gamma_k} =\sum_{k}\delta \Psi_{k}(t_j)\equiv\sum_{k}\delta \Psi_{k,j}, \label{eq:PN_exp}
    \end{split}
\end{equation}
where $\gamma_k$ are power law indices as they appear in Post-Newtonian theory ~\cite{Creighton:2011zz,Blanchet:2013haa}, and $\alpha_k$ are power law coefficients. Substituting this expression in \cref{diff_likelihood} gives
\begin{equation}
    \begin{split}
         \Rightarrow \text{ln}\mathcal{L}_o-\text{ln}\mathcal{L} =\sum_k \sum_{i,j = 0}^{N-1}&\Big[d_iC_{ij}^{-1}(-F_+h_{o,\times,j}+ F_\times h_{o,+,j})\\&+F_+ s_{o,i}C_{ij}^{-1} h_{o,\times,j}\\&-F_\times s_{o,i}C_{ij}^{-1} h_{o,+,j}\Big] \delta\Psi_{k,j}
         \\\equiv \;\sum_k P_k.
    \end{split}
\end{equation}
For a small perturbation, the two likelihood values should be nearly identical. This implies that $P_k$ should be small i.e. $\delta \Psi_{k}$ should vary only by a few phase cycles at all times,
\begin{equation}
    \delta\Psi_{k} = \delta \Psi_{k}(t) \leq 2\pi \chi \Rightarrow \delta\Psi^{\text{max}}_k=2\pi\chi\;,\label{eq:psi_max}
\end{equation}
where $\chi$ is a tunable parameter.
Using \cref{eq:PN_exp,eq:psi_max}, we can find the corresponding change in coefficient $\delta \alpha_k$:
\begin{equation}
    \begin{split}
        |\delta \alpha^{\text{max}}_k| t_*^{\gamma_k}= 2\pi\chi\Rightarrow |\delta \alpha^{\text{max}}_k| = \frac{2\pi\chi}{t_*^{\gamma_k}},
    \end{split}\label{eq:max_alpha}
\end{equation}
where for a time range $[t_\text{min},t_\text{max}]$, $t_*$ is defined as
\begin{equation}
        t_*= 
\begin{cases}
    
    |t_\text{max}|,& \text{if } \gamma_k>0\\
    |t_\text{min}|,& \text{if }\gamma_k<0
\end{cases}\label{eq:t_*}
\end{equation}

Combining \cref{eq:psi_max,eq:max_alpha,eq:t_*}, the maximum phase perturbation $\delta \Psi^\text{max}$ is written as 
\begin{equation}
    \begin{split}
        \delta \Psi^{\text{max}} = 2\pi \chi \sum_{k} \left(\frac{t}{t_*}\right)^{\gamma_k}\text{sgn}(\gamma_k).
    \end{split}\label{eq:max_phase_pertb}
\end{equation}
The term sgn$(\gamma_k)$ is included to maximize the differential phase perturbation within each bin. Finally, we impose our binning criterion, which states that within a bin $b$, the change in maximum phase perturbation must be less than a small tolerance, $\epsilon$. That is,
\begin{equation}
    \begin{split}
        |\delta\Psi^{\text{max}}(t_{f}(b)) - \delta \Psi ^{\text{max}}(t_i(b))|<\epsilon,
    \end{split}\label{eq:bin_criteria}
\end{equation}
where $t_i(b)$ and $t_f(b)$ are the initial and final times in bin $b$, or the bin edges.
The \cref{eq:max_phase_pertb,eq:bin_criteria} together define an adaptive binning scheme where $\chi$ and $\epsilon$ control the number of bins, the width of each bin and the accuracy of linear approximation for the ratio $r^{\ell m}$ within each bin.

\setlength{\tabcolsep}{3pt}
\renewcommand{\arraystretch}{1.5}
\begin{table*}[htb]
\centering
    \caption{Time taken per likelihood evaluation using time-domain heterodyning, full time-domain likelihood, frequency-domain heterodyning and full frequency-domain likelihood for various signal durations. This table also lists the number of bins required by both heterodyning methods, time taken for summary data computation in time-domain, and speedup factors achieved by heterodyning algorithm in time-domain. The time-domain methods use \texttt{IMRPhenomT}~\cite{Estelles:2020osj}, and the frequency-domain methods use \texttt{IMRPhenomXAS}~\cite{Pratten:2020fqn} waveform approximants.}
    \begin{tabular}{|
        >{\centering\arraybackslash}p{1.8cm} |
        >{\centering\arraybackslash}p{1.6cm} |
        >{\centering\arraybackslash}p{2.0cm} |
        >{\centering\arraybackslash}p{2.0cm} |
        >{\centering\arraybackslash}p{1.8cm} |
        >{\centering\arraybackslash}p{1.5cm} |
        >{\centering\arraybackslash}p{1.6cm} |
        >{\centering\arraybackslash}p{2.0cm} |
        >{\centering\arraybackslash}p{1.8cm} |}
        \hline
         \multicolumn{1}{|c|}{\begin{tabular}[c]{@{}c@{}}\textbf{ }\end{tabular}}& 
         \multicolumn{5}{c|}{\begin{tabular}[c]{@{}c@{}}\textbf{Time Domain}\end{tabular}} &
         \multicolumn{3}{c|}{\begin{tabular}[c]{@{}c@{}}\textbf{Frequency Domain}\end{tabular}} \\
        \cline{2-9} 
        \textbf{Duration of GW signal [s]} &
        \textbf{Number of bins} &
        \textbf{Summary data computation time [s]} &
        \textbf{Heterodyned Likelihood [ms]} & 
        \textbf{Full Likelihood [ms]} & 
        \textbf{Speedup factors} &
        \textbf{Number of bins} &
        \textbf{Heterodyned Likelihood [ms]} & 
        \textbf{Full Likelihood [ms]}\\ 
        \hline
        2 & 191 & \num{8.6989} & \num{2.290} & \num{42.9296} & \num{18.74655} & 120& \num{1.445}& \num{4.596}\\
        4 & 193 & \num{53.9353} & \num{2.418} & \num{89.97518} & \num{37.21057} & 121& \num{1.724} & \num{8.662}  \\ 
        8 & 302 & \num{116.1305} & \num{3.860} & \num{216.558} & \num{56.10310} & 121 & \num{1.812}& \num{14.14} \\ 
        16 & 382 &\num{561.3549} &\num{6.378} & \num{791.557}& \num{124.10740} & 121 & \num{1.801}& \num{27.35} \\ 
        32 & 476 & \num{967.96419} & \num{8.580} & \num{884.42} & \num{103.07925} & 121&\num{1.848}&\num{51.18} \\
        64 & 483 & \num{4103.1448}& \num{9.140} & \num{4163.54} & \num{455.5295} & 121& \num{1.817}&\num{105.3} \\
        128 & 486  & \num[exponent-mode=scientific]{20012.4119}& \num{9.191} & \num{3129.884} & \num{340.5379} & 121&\num{1.849}& \num{209.6} \\ 
        \hline 
    \end{tabular}
\label{Table:Speedup}
\end{table*}

We apply this binning criterion only to the inspiral part of the signal.
For the merger and ringdown parts, the bins are chosen such that they are narrow close to the merger, where the waveform evolves rapidly, and broader away from the merger, where the waveform evolves slowly. The merger and ringdown region is divided into multiple segments, each segment containing bins of constant width. The bin width in each segment increases as we move to the later ringdown part. This helps in achieving high accuracy in every part of the signal with a minimum number of bins. An illustration of time bins is shown in \cref{fig:Illustration of time bins}.

For waveforms containing subdominant modes, the bins can be chosen individually for each $(\ell, m)$ mode. By doing this, we get fewer bins for $(2,1)$ mode, but a significantly larger number of bins for $(3,3)$ or $(4,4)$ mode. This can also be seen from \cref{fig:lm_mode_ratio}, where the ratio of the $(2,1)$ modes varies slowly, while the ratio of the $(3,3)$ modes varies rapidly as a function of time.

\section{Results}\label{s1:Results}
In this section, we demonstrate both the correctness and efficiency of our formulation for heterodyning in time domain. We do so by performing parameter estimation on a set of synthetic GW signals with durations ranging from 2 seconds to 128 seconds.  While the signals analyzed in this paper can be handled in frequency domain, we use these as test cases to benchmark our implementation against frequency-domain parameter estimation methods. This validates our method before applying it to more complex scenarios such as data containing gaps, glitches or non-stationary noise, where time-domain offers significant advantages. 

We use the ``full likelihood'' or the usual Gaussian likelihood defined in \cref{eq:TD_likelihood,eq:TD_likelihood_single_detector}, as a reference in our comparisons. This likelihood is computationally intensive as it requires direct matrix multiplication between the full data, the inverse covariance matrix, and the waveform. Further, storing and inverting covariance matrices also incurs significant memory cost. We resolve this issue by using the Gohberg-Semencul~\cite {Gohberg2010} representation of the inverse of a Toeplitz matrix (see \cref{apx:Efficient_likelihood_GS_method}) to compute the full likelihood and the summary data 
\footnote{To the best of our knowledge, existing time-domain analyses in other works do not adopt this approximation, and could be sped up significantly by including it.}.

\begin{figure*}[htb]
    \begin{subfigure}[b]{0.47\textwidth}
        \centering
        \includegraphics[width=1.0\linewidth]{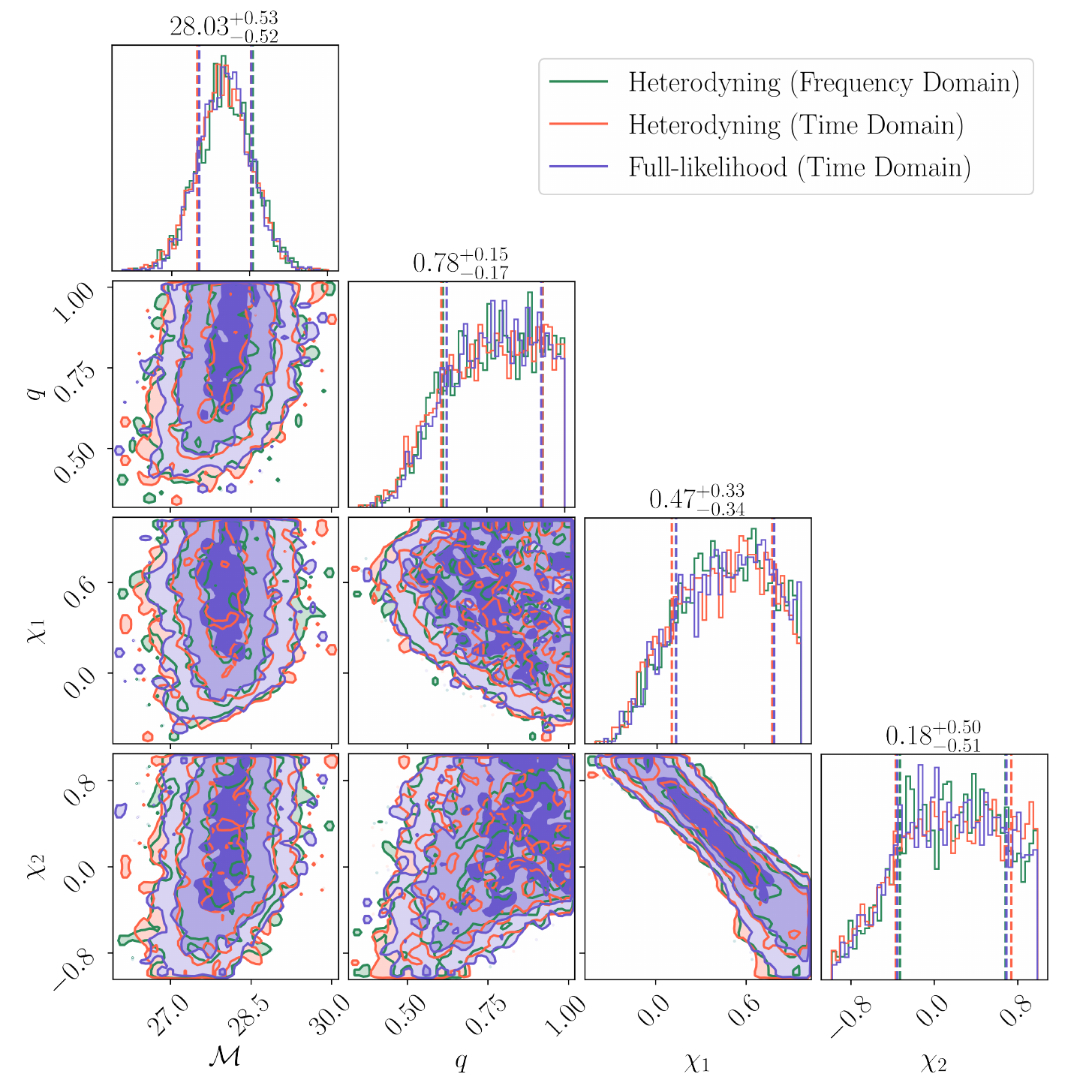} 
        \caption{}
        \label{fig:2_second_example_intrinsic}
    \end{subfigure}
    \begin{subfigure}[b]{0.47\textwidth}
        \centering
        \includegraphics[width=1.0\linewidth]{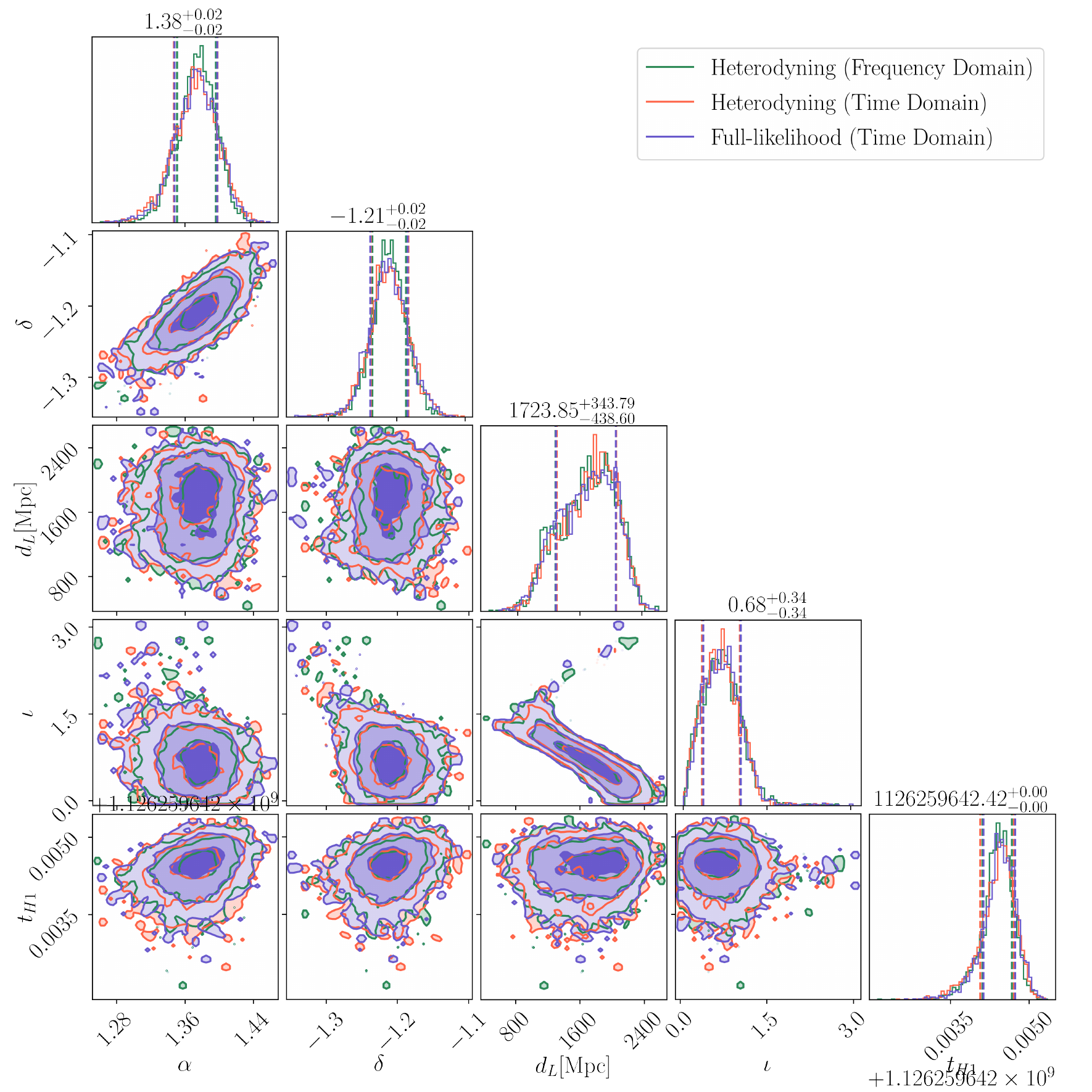} 
        \caption{}
        \label{fig:2_second_example_extrinsic}
    \end{subfigure}
    \caption{Comparison between posterior probability distribution of intrinsic (left panel) and extrinsic (right panel) parameters of a 2-second long GW signal, estimated using time-domain heterodyning, frequency domain heterodyning~\cite{Krishna:2023bug} and full time-domain likelihood method. The time-domain analyses use \texttt{IMRPhenomT} and frequency domain analysis uses \texttt{IMRPhenomXAS}~\cite{Pratten:2020fqn} waveform approximant. All three methods produce identical posteriors. The maximum JS divergence between both time-domain methods is 0.00191 across all sampled parameters, while the divergence between frequency-domain and time-domain heterodyning is 0.00397. The time-domain heterodyning achieves this accuracy while being 19$\times$ faster than the full likelihood method. The time required for a single likelihood evaluation, and the injection parameters are listed in \cref{Table:Speedup} and \cref{Table:Injection_parameters}, respectively.}
    \label{fig:2_second_example}
\end{figure*}
In the following subsections, we present a comparison between the full likelihood method and heterodyning in time-domain, the speedup factors achieved by heterodyning, and p-p plots for BBH and BNS injections. All parameter estimation analyses in this paper are carried out using the \texttt{Dynesty} nested sampler~\cite{Speagle_2020} implemented in \texttt{Bilby}~\cite{Ashton_2019}. The covariance matrices in each case are constructed using the \texttt{aLIGO\_O4high} sensitivity curve for the LIGO detectors~\cite{T2000012,abbott2020prospects} and the \texttt{BNS Optimised} curve for Virgo~\cite{P1200087,abbott2020prospects}, both availed via the \texttt{Bilby} package\footnote{\href{https://github.com/bilby-dev/bilby/tree/main/bilby/gw/detector/noise\_curves}{https://github.com/bilby-dev/bilby/tree/main/bilby/gw\\/detector/noise\_curves}}. We use \texttt{IMRPhenomT}~\cite {Estelles:2020osj} waveform approximant with a sampling rate, $f_s$, of 4096 Hz in each analysis. 

A key requirement of our method is the generation of waveform modes at non-uniformly spaced bin edges. Therefore, we modified the source code for \texttt{IMRPhenomT}/\texttt{IMRPhenomTHM}~\cite{Estelles:2020twz} waveform models implemented in \texttt{LALSuite}~\cite{lalsuite}\footnote{ The modified version of code can be found \href{https://github.com/neha2023sharma/lalsuite/blob/a13e410022d5db89c64983c2bf9c1c1da54f7cdb/lalsimulation/lib/LALSimIMRPhenomTHM.c\#L794}{here}.}. The modified version allows the generation of waveform modes at user-specified time arrays, which helps achieve maximum efficiency while downsampling the waveform modes. Generally, the waveforms are generated at the geocenter and then shifted to account for detector time delays. In our implementation, however, we shift the waveform (or modes) during the generation process by passing a time array $T-t_{\text{delay}}$ directly to the waveform generator. 

\begin{figure*}[htb]
     \begin{subfigure}[b]{0.47\textwidth}
         \centering
        \includegraphics[width=1.0\linewidth]{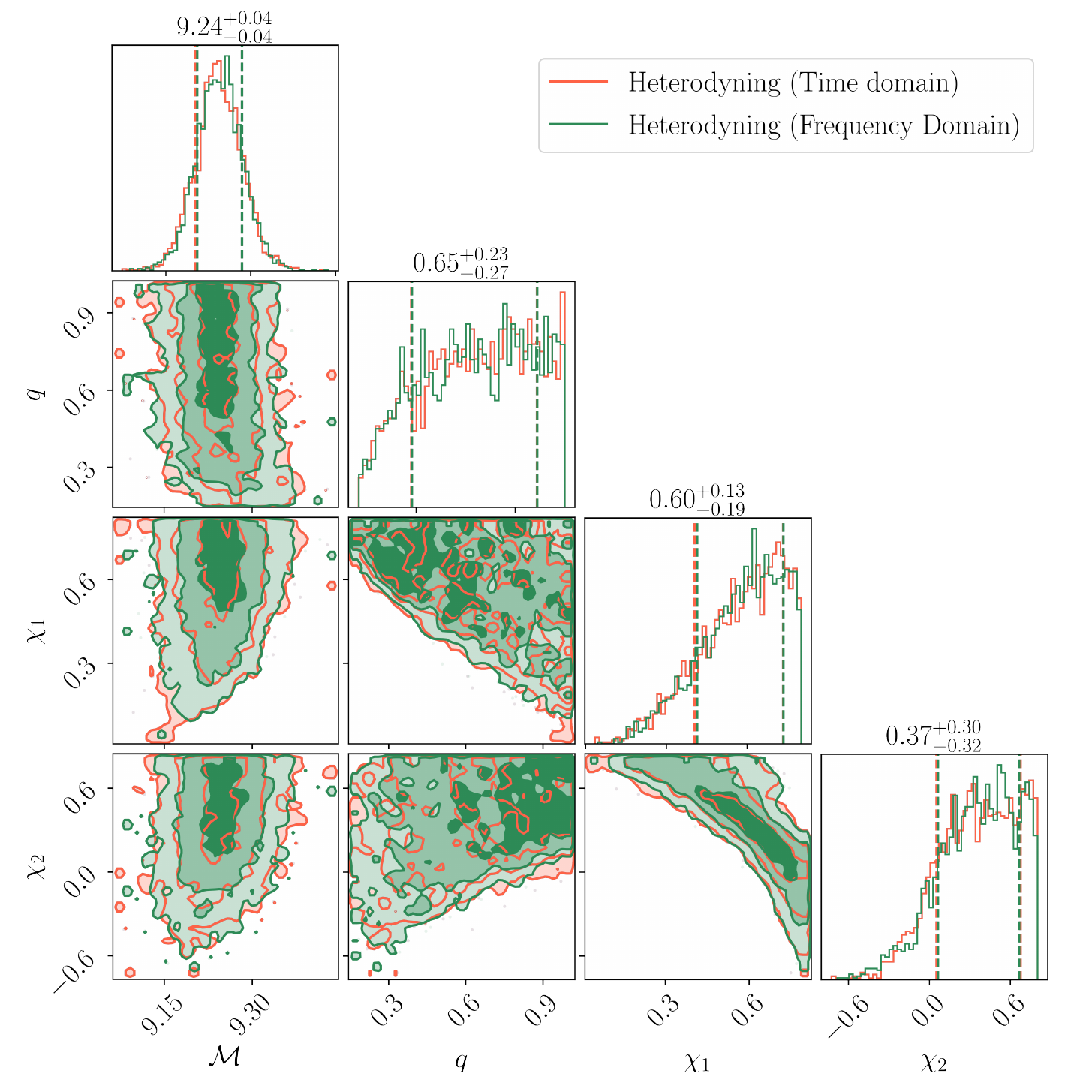} 
        \caption{}
        \label{fig:16_second_example_intrinsic}
    \end{subfigure}
    \begin{subfigure}[b]{0.47\textwidth}
         \centering
        \includegraphics[width=1.0\linewidth]{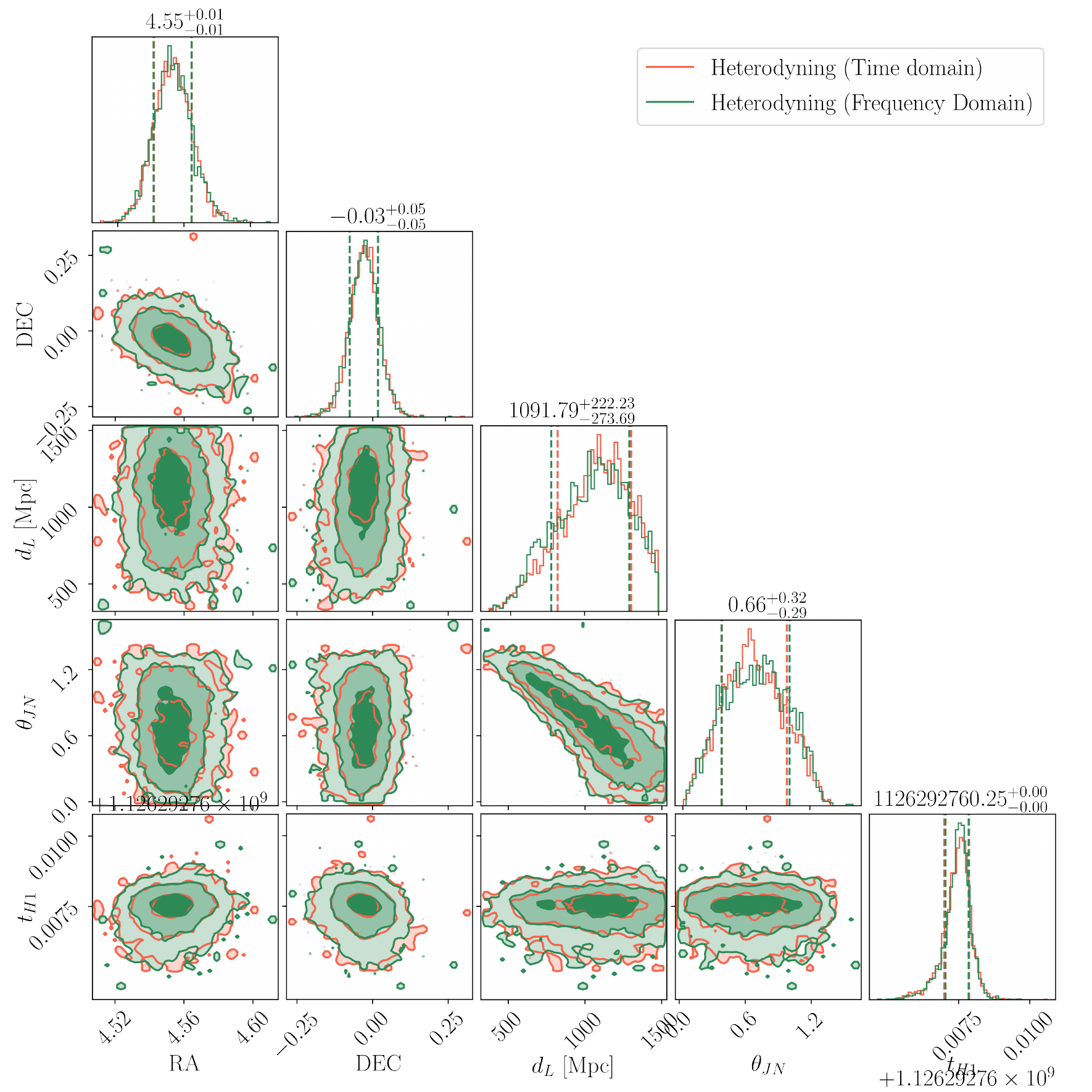} 
        \caption{}
        \label{fig:16_second_example_extrinsic}
    \end{subfigure} 
    \caption{Comparison between posterior probability distribution of intrinsic (left panel) and extrinsic (right panel) parameters of a 16-second long GW signal, estimated using time-domain heterodyning and frequency domain heterodyning~\cite{Krishna:2023bug}. The time-domain analysis uses \texttt{IMRPhenomT} and frequency domain analysis uses \texttt{IMRPhenomXAS}~\cite{Pratten:2020fqn} waveform approximant. Both methods produce identical posteriors, with maximum JS divergence of 0.00196 across all sampled parameters. The time required for a single likelihood evaluation, and the injection parameters are listed in \cref{Table:Speedup} and \cref{Table:Injection_parameters}, respectively.}
    \label{fig:16_second_example}
\end{figure*}
\begin{figure*}[htb]
     \begin{subfigure}[b]{0.47\textwidth}
         \centering
        \includegraphics[width=1.0\linewidth]{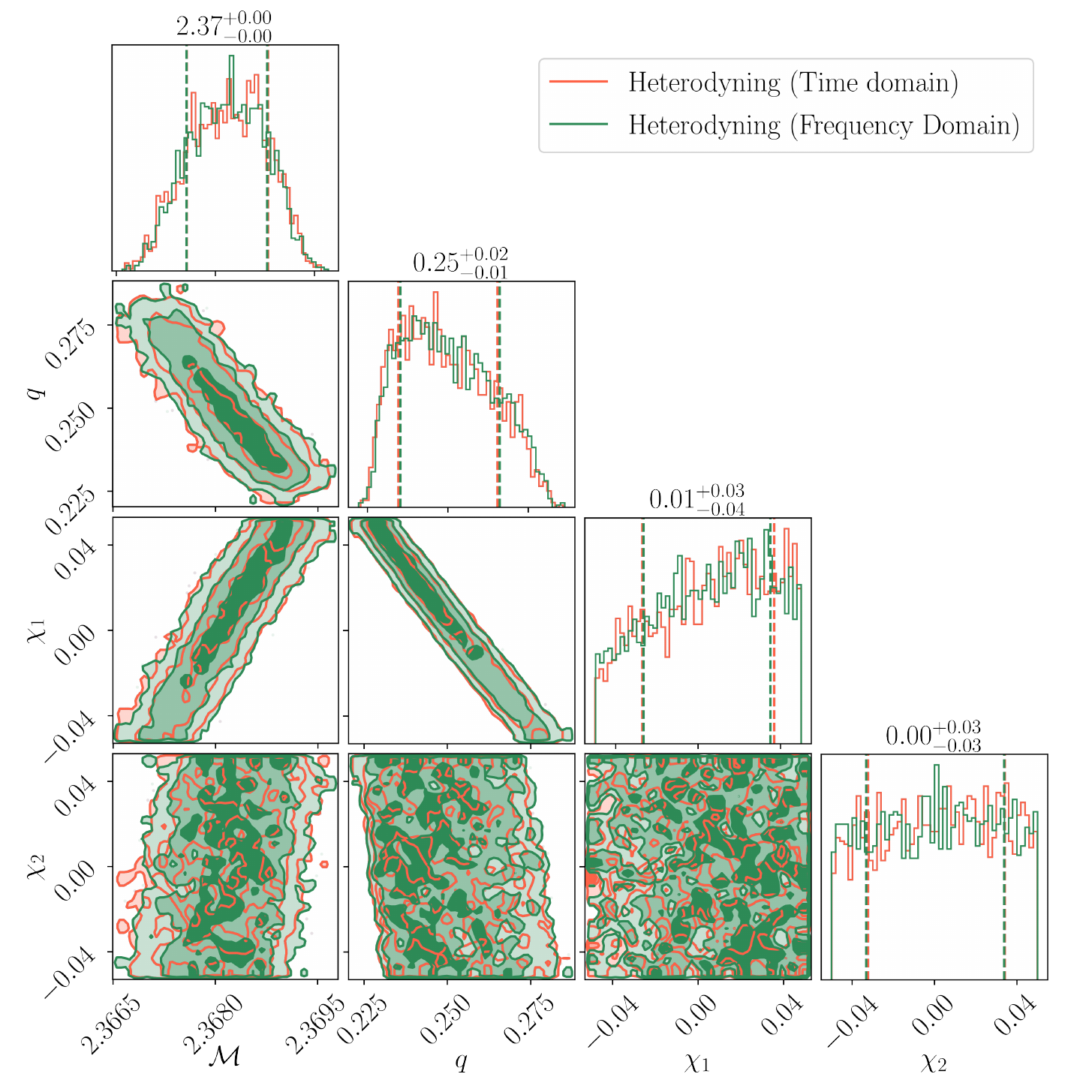} 
        \caption{}
        \label{fig:128_second_example_intrinsic}
    \end{subfigure}
    \begin{subfigure}[b]{0.47\textwidth}
         \centering
        \includegraphics[width=1.0\linewidth]{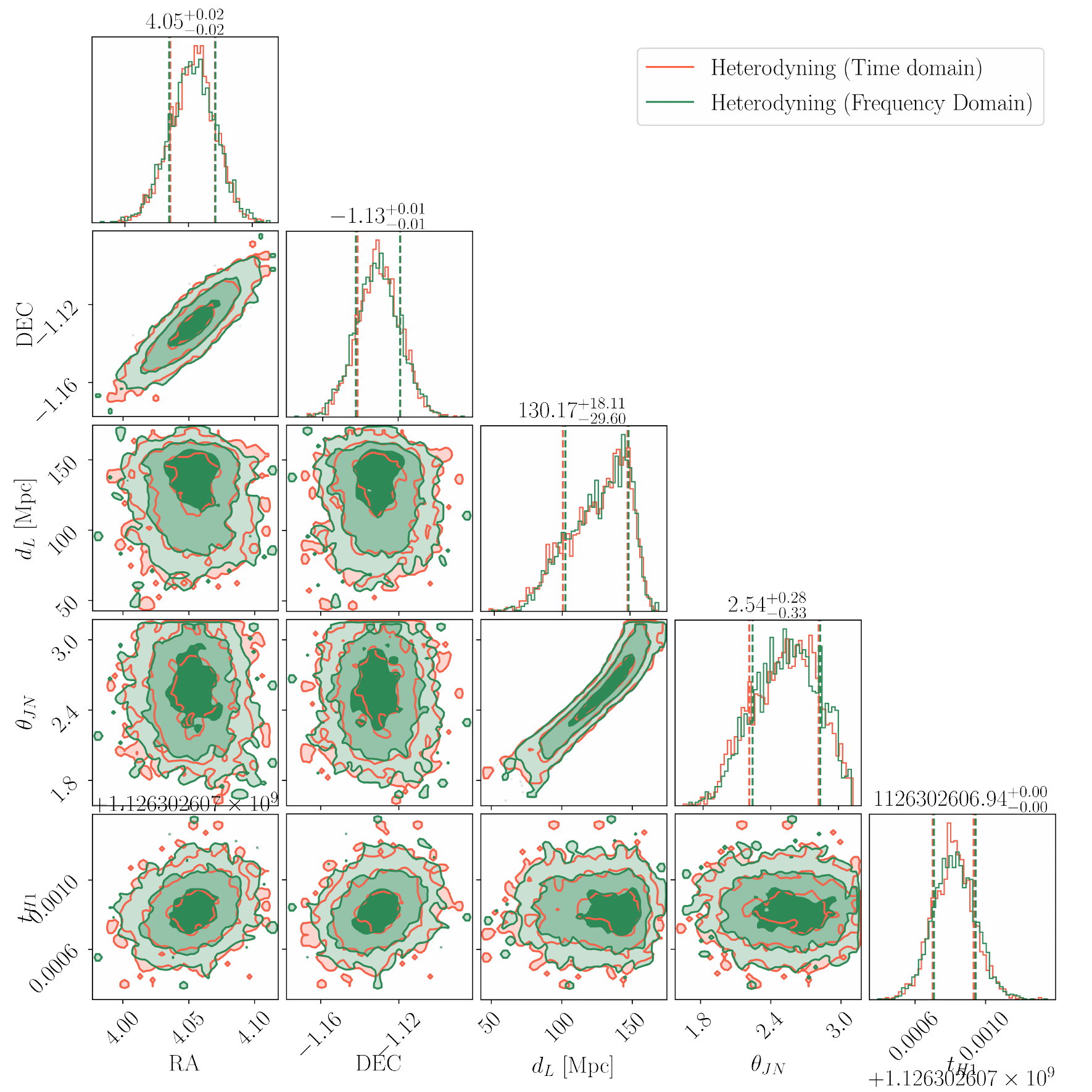} 
        \caption{}
        \label{fig:128_second_example_extrinsic}
    \end{subfigure}
    \caption{Similar to \cref{fig:16_second_example}, we show comparison between the posterior probability distribution of intrinsic (left panel) and extrinsic (right panel) parameters of a 128-second aligned-spin BBH signal, recovered using time-domain and frequency-domain heterodyning. Both methods produce identical posteriors, with maximum JS divergence of 0.00133 across all sampled parameters.}
    \label{fig:128_second_example}
\end{figure*}

%
\setlength{\tabcolsep}{10pt}
\renewcommand{\arraystretch}{1.5}
\begin{table*}[htb]
\caption{Injection parameters used in \cref{fig:2_second_example,fig:16_second_example,fig:128_second_example}}
\centering
    \begin{tabular}{|c|c|c|c|}
          \hline
          \textbf{Injection Parameters} & \textbf{Figure~\ref{fig:2_second_example}} & \textbf{Figure~\ref{fig:16_second_example}} & \textbf{Figure~\ref{fig:128_second_example}} \\ 
          \hline 
          Detector frame chirp mass $\mathcal{M}[M_\odot]$ & \num{28.0955} & \num{9.24566486081656} & \num{2.368726082867532}  \\ 
          Mass ratio $q$ & \num{0.8055} & \num{0.6972825312736546} & \num{0.23665589518271324} \\ 
          Primary spin $\chi_1$ & \num{0.4} & \num{0.5989593371623323}& \num{0.026325207825307853}\\
          Secondary spin $\chi_2$ & \num{0.3} & \num{0.331142670582826}  &\num{0.03565162871375553}\\
          Luminosity distance $d_L$ [Mpc] & \num{2000.0} & \num{559.1415145588842} & \num{142.62506256493438} \\ 
          Right ascension $\alpha$ & \num{1.375} & \num{4.548787435158331} &  \num{4.053024565064559} \\ 
          Declination $\delta$ & \num{-1.2108} & \num{-0.02319639437997357} & \num{-1.1313318063665236} \\ 
          Polarization angle $\psi$ & \num{2.659} & \num{2.589244795102876} & \num{2.2624402074285253} \\ 
          Inclination angle $\iota$ & \num{0.4} &  \num{1.1559239659720089} & \num{2.7466268917518435} \\ 
          Reference phase $\phi$ & \num{1.3} & \num{4.500854280282311} & \num{4.78228278738645}\\
          Arrival time at H1 detector  $t_{\mathrm{H1}}$ & 1126259642.42 & 1126292760.25 & 1126302606.94 \\
          \hline
          Duration & 2 seconds & 16 seconds & 128 seconds \\
          Network SNR & 18.35 & 9.39 & 30.51\\
          \hline
    \end{tabular}
\label{Table:Injection_parameters}
\end{table*}

\begin{figure*}[htb]
     \begin{subfigure}[b]{0.47\textwidth}
         \centering
         \includegraphics[width=\linewidth]{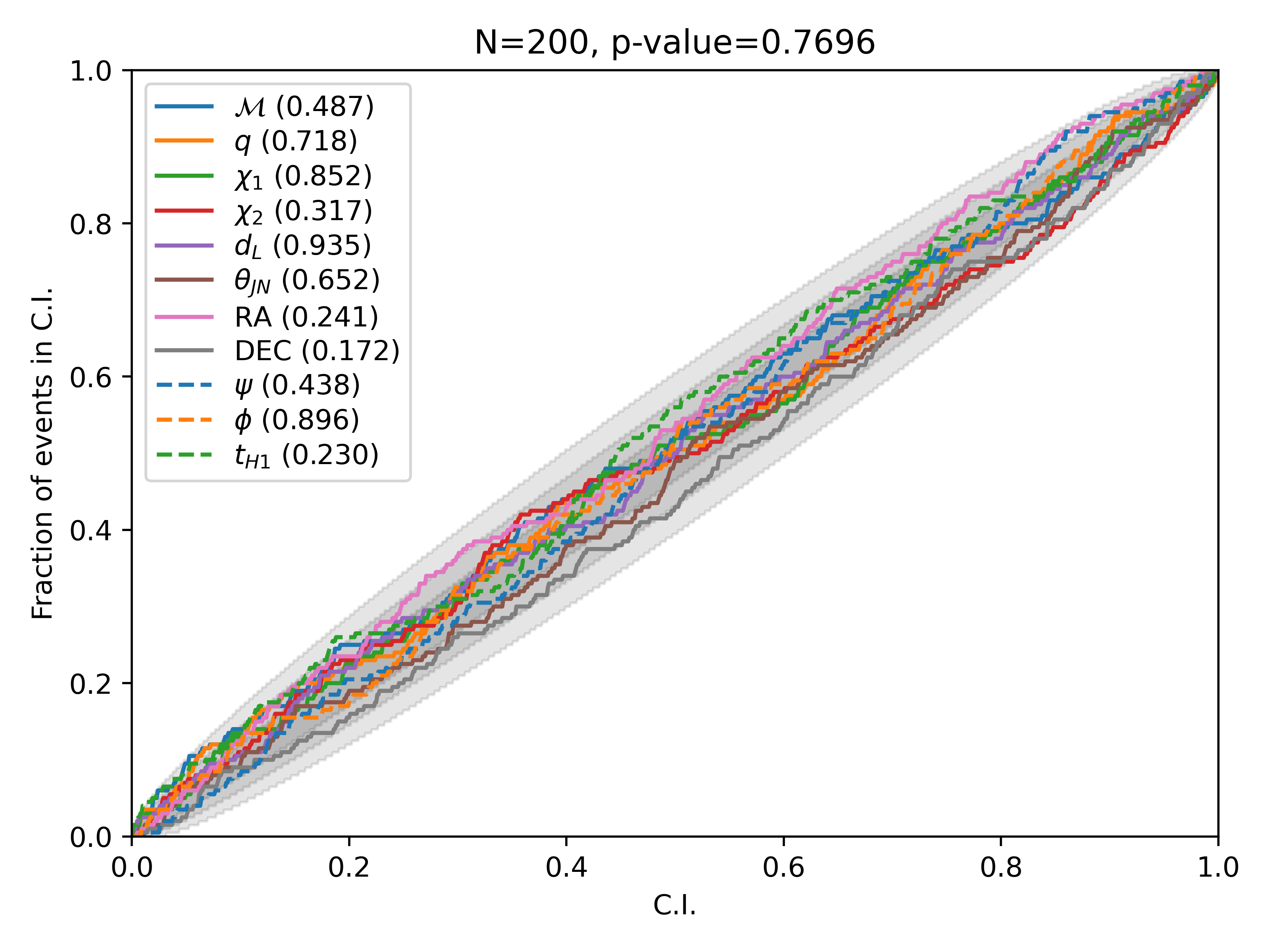}
         \caption{}
         \label{fig:8_sec_pp_plot}
     \end{subfigure}
     \hfill
     \begin{subfigure}[b]{0.47\textwidth}
         \centering
         \includegraphics[width=\linewidth]{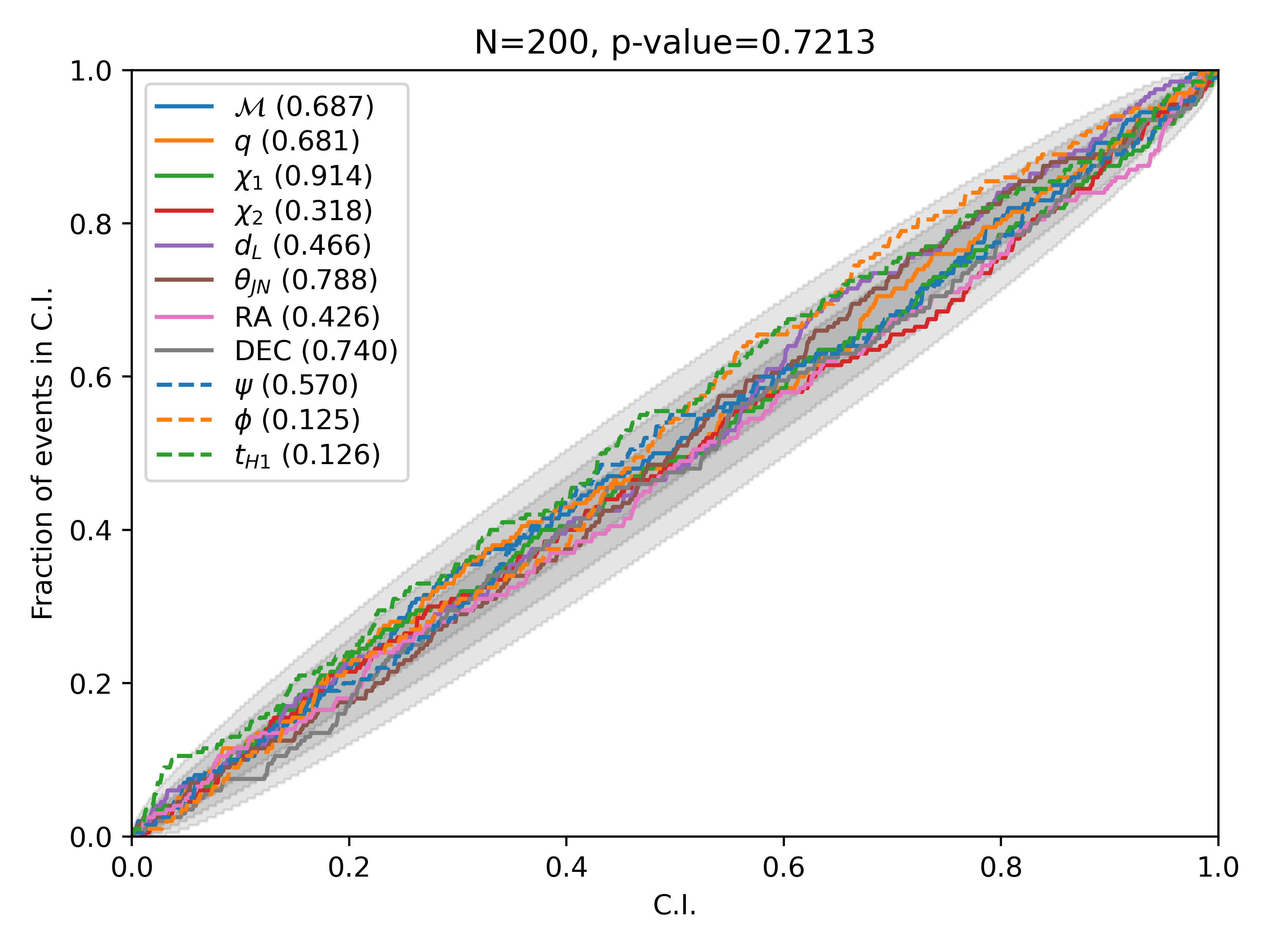}
         \caption{}
         \label{fig:128_sec_pp_plot}
     \end{subfigure}
     \caption{Percentile-Percentile (p-p) plots for (a) 8-second BBH, (b) 128-second BNS injections in Gaussian noise. The x-axis represents the credible interval, and y-axis shows the fraction of events for which the injected value lies within that interval. The grey shaded area represents statistical errors or fluctuations due to a finite number of injections. If curves for various parameters lie within this shaded region, it implies that our analysis is free from any significant bias. The numbers in parentheses are p-values obtained from the Kolmogorov-Smirnov test. A low p-value (<0.05) indicates a bias, and a high p-value means that the posteriors are consistent with the true value. The priors used to simulate injections and recover posteriors are listed in \cref{Table:Priors_for_PP_plots}.}
\end{figure*}

\setlength{\tabcolsep}{10pt}
\renewcommand{\arraystretch}{1.5}
\begin{table*}[htb]
\centering
    \caption{Prior distributions used for simulating BBH and BNS injections, as well as for recovery.}
    \begin{tabular}{|c|c|c|}
        \hline
        \textbf{Parameter} & \textbf{BBH Prior} & \textbf{BNS Prior} \\
        \hline
        Chirp mass $\mathcal{M}[M_\odot]$ & Uniform $[7.93,14.75]$ & Uniform $[1.42,2.60]$ \\
        Mass ratio $q$ & Uniform $[1/6,1]$ & Uniform $[1/6,1]$ \\
        Aligned spins $\chi_1, \chi_2$ & Uniform $[-0.8,0.8]$ & Uniform $[-0.05,0.05]$ \\
        Luminosity distance $\,D_L$ [Mpc] & $D_L^2$ prior, $D_L\in [50,500]$ & $D_L^2$ prior, $D_L\in [10,\,150]$ \\
        Right ascension $\alpha$ & Uniform $[0,2\pi]$, periodic & Uniform $[0,2\pi]$, periodic \\
        Declination $\delta$ & Cosine prior & Cosine prior \\
        Inclination angle $\iota$ & Sine prior & Sine prior \\
        Polarization angle $\psi$ & Uniform $[0,\pi]$, periodic & Uniform $[0,\pi]$, periodic \\
        Reference phase $\phi$ & Uniform $[0,2\pi]$, periodic & Uniform $[0,2\pi]$, periodic \\
        \hline 
    \end{tabular}
\label{Table:Priors_for_PP_plots}
\end{table*}

\subsection{Illustrative Example}\label{s2:posterior_comparison}

We begin with an illustrative Bayesian analysis of a 2-second long GW signal and compare the posterior probability distributions of its parameters obtained using the heterodyning\footnote{Throughout this paper, we refer to time-domain heterodyning and full time-domain likelihood simply as ``heterodyning'' and ``full likelihood'' respectively. We use a prefix ``frequency-domain'' when discussing frequency domain methods.} formulation described in \cref{s2:Relative_binning,s2:Summary_data,s2:Binning_criteria} with those obtained using the full likelihood. For completeness, we also compare both time-domain methods against frequency-domain heterodyning implemented in \texttt{Bilby}~\cite{Ashton_2019,Krishna:2023bug}. As analyses using the full likelihood become prohibitively expensive for longer signals, we only compare heterodyning in the time domain to analyses in the frequency domain. 

For this comparison, we inject a 2 second-long aligned-spin BBH signal into zero-noise realization of a network of LIGO, and Virgo detectors, resulting in a network signal-to-noise ratio (SNR) of 18.35. We sample all 11 parameters using all methods. The heterodyning analysis utilizes 191 bins with fiducial parameters set equal to injection parameters.

Figures \ref{fig:2_second_example_intrinsic} and \ref{fig:2_second_example_extrinsic} demonstrate that the posteriors for intrinsic and extrinsic parameters recovered using both time-domain methods and frequency-domain heterodyning are in excellent agreement. The time-domain heterodyning accurately captures various parameter degeneracies, such as degeneracies between component spins ($\chi_1, \chi_2$), and between luminosity distance ($d_L$) and inclination angle ($\theta_{JN}$). To quantify this agreement, we compute Jensen-Shannon(JS) divergence~\cite{lin2002divergence} for each sampled parameter. The largest JS divergence across all sampled parameters and all methods is 0.00397, which is significantly lower than the threshold of 0.06, also used in~\cite{Krishna:2023bug}. This concludes that the posteriors constructed using heterodyning, full likelihood method and frequency-domain heterodyning are statistically identical.

Beyond its accuracy, heterodyning also demonstrates remarkable computational efficiency. On single CPU core, full likelihood takes $43$ ms, but heterodyning only takes $2.3$ ms for a single likelihood evaluation, which is approximately 19 times faster (see \cref{Table:Speedup}). When parallelized over 16 CPU cores, heterodyning finishes the analysis in just 1 hour and 40 minutes, while full likelihood requires 32 hours. This speedup increases dramatically for BNS systems and signals in next-generation detectors.


\subsection{Large-scale Injection study}
To demonstrate the scalability of our method, we extend our analysis to longer GW signals with durations of 16 and 128 seconds. As in previous section, we use zero-noise realization and inject signals in a network of LIGO and Virgo detectors. The injection parameters and corresponding network SNRs for each case are listed in \cref{Table:Injection_parameters}. 

As shown in \cref{fig:16_second_example,fig:128_second_example}, the recovered posteriors are consistent with the injection parameters and identical to those estimated using frequency-domain heterodyning. The maximum JS divergence between heterodyning in time-domain and frequency-domain is 0.00196 for 16-second long signal and 0.00133 for 128-second long signal. The time taken for a single likelihood evaluation for a 16 second-long GW signal is 6.4 ms with heterodyning, which is approximately 124 times faster than the full likelihood method. This speedup increases to a factor of 341 for 128 second-long signal, finishing a full analysis in just a few hours when parallelised over 16 CPU cores. 

We further test our method across the parameter space of BBH and BNS systems. For this purpose, we inject 200 BBH signals with durations of 8 seconds, and chirp mass ranging from 7.93 $M_\odot$ to 14.75 $M_\odot$, into simulated Gaussian noise and make p-p plots of the recoveries. We repeat the same for 200 BNS injections with durations of 128 seconds, and chirp mass ranging between 1.42 $M_\odot$ and 2.60 $M_\odot$. All signals are injected into a network of LIGO and Virgo detectors. The priors used to simulate injections and to recover posteriors are listed in \cref{Table:Priors_for_PP_plots}. 

In a p-p plot, the x-axis represents a credible interval of the recovered posteriors, and the y-axis shows the fraction of events for which injected (true) value lies within that credible interval. For an unbiased analysis, a credible interval of $90\%$ (or $p\%$) should contain the true value $90\%$ (or $p\%$) of the time. So, the curves for various recovered parameters should lie along the diagonal. But due to a finite number of injections, these curves may deviate from the diagonal. These deviations or statistical fluctuations are shown by grey bands in \cref{fig:8_sec_pp_plot,fig:128_sec_pp_plot}, which correspond to $1\sigma, \;2\sigma $ and $3\sigma$ confidence intervals of a uniform distribution.
We expect $68.3\%$, $95.4\%$ and $99.7\%$ of the curves for various parameters to lie within these bands if our analysis is free from any significant bias. These grey bands shrink as the number of simulated signals increases. 

\cref{fig:8_sec_pp_plot,fig:128_sec_pp_plot} show that our recoveries are consistent with the error regions. The p-value of each sampled parameter is greater than $ 0.05$, which provides strong evidence against the presence of any bias. We conclude that our method recovers unbiased and reliable posteriors for both BBH and BNS injections.

While we successfully recovered the posteriors for the majority of injection parameters, a small number ($8\%-12\%$) of injections did not converge to the correct value. This happens when the waveform amplitude in the ringdown regime approaches zero, causing the ratio of modes to diverge. 
We find that this issue can be mitigated either by increasing the number of bins (or reducing $\epsilon$) in the ringdown regime or by using tighter priors on the arrival times. 


\subsection{Relative Speedup}
\begin{figure*}[htb]
    \centering
    \includegraphics[width=1.0\linewidth]{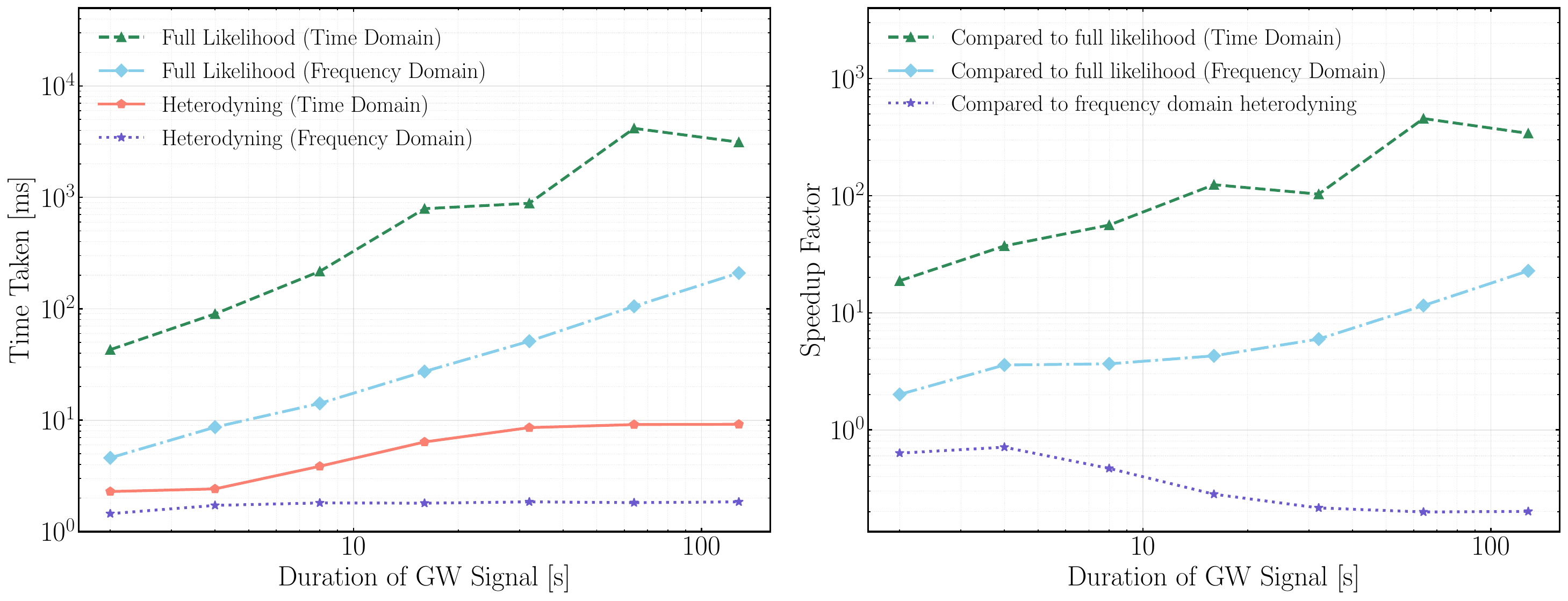}
    \caption{(a) Time taken for a single likelihood evaluation (on one CPU core) using four methods: full time-domain likelihood, full frequency-domain likelihood, time-domain heterodyning and frequency-domain heterodyning. (b) The scaling of speedup factors achieved by time-domain heterodyning against the other three methods. The time-domain heterodyning achieves significant speedup against time domain and frequency domain full likelihoods, but it is computationally more expensive than the frequency-domain heterodyning.}
    \label{fig:Comparison_time_frequency_domain}
\end{figure*}

The computational cost of the heterodyned time-domain likelihood is $\sim 20$--$300\times$ lower than that of the full time-domain likelihood (see Table~\ref{Table:Speedup}), and is comparable to the cost of the heterodyned frequency-domain likelihood~\cite{cornish2021heterodyned, zackay2018relative, Krishna:2023bug}. This speedup is achieved through the following optimizations: 
\begin{enumerate}
    \item Faster waveform generation: The time required to generate a single waveform can drastically increase with signal duration. Instead of generating waveforms at $N = T\times f_s$ sample points, the heterodyning generates waveform modes only at the bin edges, which typically amounts to only a few hundred sample points. 
    Moreover, our modified version of \texttt{IMRPhenomT} enables direct downsampling of waveforms without any computational overhead of interpolation. Consequently, the waveform generation time scales linearly with the number of bins, making it extremely efficient, especially when the number of bins is small ($\sim \mathcal{O}(10^2)$).
    \item Faster Likelihood computation: The primary computational cost in a likelihood evaluation is the matrix multiplication between the signal and the inverse of the covariance matrix, $\sum_{j}C^{-1}_{ij}h_{j}$. 
    In heterodyning, however, the likelihood is reformulated such that the terms containing covariance matrix are computed in advance. Due to this, the likelihood depends only on the bin edges and requires only $\mathcal{O}(N_\text{bins})$ operations. 
\end{enumerate}

Both of these advantages become increasingly significant for longer GW signals. As the signal duration $T$ increases, number of sample points $N$ increases linearly with it, whereas the total number of bins scales much more slowly. For example, a 128-second GW signal sampled at 4096 Hz would contain approximately $\sim 5\times 10^5$ data points, but can be analyzed with only $\sim 500$ time bins using heterodyning.

We also compare the time-domain heterodyning against full time-domain likelihood, full frequency-domain likelihood, and frequency-domain heterodyning implemented in \texttt{Bilby}~\cite{Ashton_2019, Krishna:2023bug}. We use \texttt{IMRPhenomT} waveform approximant for time-domain methods and \texttt{IMRPhenomXAS} waveform approximant for frequency-domain methods. \cref{Table:Speedup} shows the total number of bins required for both heterodyning algorithms, as well as the single likelihood evaluation time using all four methods. We note that both heterodyning approaches require different number of bins; the time domain requires a slightly larger number of bins to achieve the desired accuracy. In \cref{fig:Comparison_time_frequency_domain}, we show the scaling behaviour of single likelihood evaluation time using all methods and speedup factors achieved by time-domain heterodyning against the other three methods.


\section{Conclusion and Future Prospects}
In this work, we presented a highly efficient and accurate framework for parameter estimation of gravitational waves using heterodyning in the time domain. Our method dramatically accelerates the likelihood computation in two key steps. First, we downsample the waveform modes to just a few hundred sample points. Second, we precompute the computationally expensive terms involving the covariance matrix and store them as summary data. The likelihood then depends only on the summary data and the ratio of waveform modes evaluated at bin edges. This significantly reduces the likelihood evaluation time, and this computational cost scales linearly with the total number of bins. 

We also derive an adaptive binning criterion to achieve a desired accuracy while simultaneously minimizing the number of bins. For the inspiral phase, we derived a binning criterion which states that the maximum phase difference between sampled and fiducial waveform modes should not exceed a small number $\epsilon$ over a time bin. This results in wider bins during early inspiral and narrower bins close to the merger. For merger and ringdown phases, we choose uniformly spaced bins that are densely spaced in early ringdown and are wider during the late ringdown part of GW signals. 

We further validate the versatility and robustness of the scheme by testing it across the temporal and mass regimes of compact binary coalescences. We analyze populations of BBH and BNS injections in the network of LIGO-Virgo detectors. Through these we confirmed that we accurately recover parameters for both short-duration BBH signals and long-duration BNS events, with no significant bias. 

Heterodyning dramatically improves the computational efficiency of time-domain parameter estimation, and more importantly makes time-domain analyses of long BNS signals practically tractable. As demonstrated in \cref{fig:Comparison_time_frequency_domain}, for a 128-second long signal, our method is at least $340 \times$ faster than standard Bayesian analysis described in \cref{s2:bayesian_pe}. As a result, heterodyning can finish the analysis of 128-second BNS signals in just a few hours, rather than weeks or months required by standard Bayesian method.

Our method is the most effective with models from the phenomenological waveform family, as they can be easily modified to generate waveforms at arbitrary times. During the course of this work, we also performed a limited number of analyses with numerical relativity surrogate models~\cite{Blackman:2017pcm,Varma:2018mmi, Varma:2019csw}. These models support waveform generation at arbitrary time points; however, each waveform evaluation incurs a fixed cost to set up the time interpolant as well as a variable cost to evaluate that interpolant that scales with the number of time points. Consequently, with current surrogate models, the speed-up would saturate as a function of signal duration. Future improvements in reducing the aforementioned fixed cost in the surrogate waveform generation can lead to speed-up similar to those seen for phenomenological models. 
In contrast, the Effective One-Body (EOB) models~\cite{Cotesta:2018fcv, Nagar:2018zoe,Ossokine:2020kjp,Ramos-Buades:2021adz,Pompili:2023tna, Ramos-Buades:2023ehm,Albanesi:2025txj} explicitly evolve differential equations numerically, which do not allow waveform generation at arbitrary times. In order to use the EOB models as implemented in \texttt{LALSuite}~\cite{lalsuite}, we have to generate waveforms on a dense time grid and then downsample to the bin edges. While we would lose any speedup due to the downsampling of the number of time points that heterodyning enables, we would still achieve a speedup due to the pre-calculation of terms involving the covariance matrix. Alternatively, this computation could in-principle be accelerated by evaluating the EOB waveform directly on the sparse time grid produced by the ODE solver, subsequently interpolating it, and then evaluating the interpolant on heterodyned time array~\footnote{We thank Lorenzo Pompili for drawing our attention to this point.}.

In this paper, we focused on the simplest analyses, using only dominant waveform modes and aligned spin systems. Looking ahead, we plan to extend this framework to handle the complexity of future observations. While this paper focused only on dominant waveform modes, heterodyning formalism can also be applied to subdominant waveform modes as well. We provide the complete theoretical framework for a mode-by-mode analysis in \cref{apx:Summary_data_subdominant_modes}.
The binning criteria described in \cref{s2:Binning_criteria} can also be applied mode-by-mode. For instance, slowly varying mode (2,1) can be approximated with fewer number of bins, while (3,3) mode requires a higher number of bins. 

The dramatic speedup achieved by heterodyning also makes it a promising tool for next generation detectors, where we must account for Earth's rotation by using time varying antenna pattern. Our framework can directly incorporate these time dependent antenna patterns with negligible increase in computational cost, as these functions need to be evaluated at bin edges rather than full time grid. Lastly, we also aim to validate our method in presence of non-stationary gaussian noise. As discussed earlier, the framework developed in this paper can incorporate any covariance matrix, as these matrices are completely absorbed into the summary data calculation. Thus, we can employ non-stationary noise models without adding any computational cost during sampling.

\section{Acknowledgements}

We thank members of the Astrophysics and Relativity group at ICTS for valuable discussions and input. We are grateful to Simona Miller for a meticulous reading of our draft. We also thank Xiao-Xiao Kou and Lorenzo Pompili for helpful comments.

This work was supported by the Department of Atomic Energy, Government of India, under project number RTI4001. A.V. acknowledges support from the Natural Sciences and Engineering Research Council of Canada (NSERC) (funding reference number 568580). P.K. acknowledges support by the Ashok and Gita Vaish Early Career Faculty Fellowship at the International Centre for Theoretical Sciences.

\bibliography{main}
\onecolumngrid
\appendix
\section{Summary data for subdominant modes}\label{apx:Summary_data_subdominant_modes}
In this section, we present a detailed derivation of likelihood (see \cref{eq:TD_likelihood}) in terms of \textit{summary data} and ratio of modes $r^{\ell m}$, for general $(\ell,m)$. We begin with the first term in \cref{eq:decomposed_likelihood}:
\begin{equation}
    \begin{split}
        L_1(d,s) &= L_1(d,F_+h_++F_\times h_\times) = F_+ L_1(d,h_+) + F_\times L_1(d,h_\times)\\
        &= F_+ L_1(d,\mathfrak{Re}(\mathcal{H})) + F_\times L_1(d,-\mathfrak{Im}(\mathcal{H}))\\
        &= F_+ \mathfrak{Re}[L_1(d,\mathcal{H})] - F_\times \mathfrak{Im}[L_1(d,\mathcal{H})],
    \end{split}\label{eq:3}
\end{equation}
where $L_1(d,\mathcal{H})$ is written in terms of waveform modes $h^{\ell m}$ as
\begin{equation}
    \begin{split}
        L_1(d,\mathcal{H}) = \sum_{\ell  = 2}^{\infty}\sum_{m = -\ell }^{\ell }\;_{-2}Y^{\ell m} L_1(d,h^{\ell m})&= \sum_{\ell =2}^{\infty} \sum_{m>0}^{\ell} (\;_{-2}Y^{\ell m} L_1(d,h^{\ell m}) + \;_{-2}Y^{\ell ,-m}L_1(d,h^{\ell ,-m}))\\
        &= \sum_{\ell =2}^{\infty} \sum_{m>0}^{\ell} (\;_{-2}Y^{\ell m} L_1(d,h^{\ell m}) + \;_{-2}Y^{\ell ,-m}(-1)^\ell  L_1(d,h^{*\; lm}))\\
        &= \sum_{\ell =2}^{\infty} \sum_{m>0}^{\ell}(\;_{-2}Y^{\ell m} L_1(d,h^{\ell m}) + \;_{-2}Y^{\ell ,-m}(-1)^\ell  L_1(d,h^{\ell m})^*).
    \end{split}\label{eq:4}
\end{equation}
Substituting this expression back into \cref{eq:3}, we obtain $L_1(d,s)$ in terms of spin weighted spherical harmonics $_{-2}Y^{\ell m}$  and waveform modes $h^{\ell m}$:
\begin{equation}
    \begin{split}
        L_1(d,s) = \;&F_+ \mathfrak{Re}\left[\sum_{\ell=2}^{\infty} \sum_{m>0}^\ell \left(\;_{-2}Y^{\ell m} L_1(d,h^{\ell m}) + \;_{-2}Y^{\ell ,-m}(-1)^\ell L_1(d,h^{\ell m})^*\right)\right]\\& - F_\times \mathfrak{Im}\left[\sum_{\ell =2}^{\infty} \sum_{m>0}^\ell \left(\;_{-2}Y^{\ell m} L_1(d,h^{\ell m}) + \;_{-2}Y^{\ell ,-m}(-1)^\ell  L_1(d,h^{\ell m})^*\right)\right].
    \end{split} \label{eq:L_d_s}
\end{equation}
The term $L_1(d,h^{\ell m})$ is further expressed in terms of fiducial modes $h^{\ell m}_o$ and ratio of waveform modes $r^{\ell m}$ as
\begin{equation}
    \begin{split}
        L_1(d,h^{\ell m}) &= \sum_{ij} d_i C_{ij}^{-1} h^{\ell m}_j = \sum_{b}\sum_{j\in b}\sum_{i = 0}^N d_i C_{ij}^{-1} h^{\ell m}_j\\
        &=  \sum_{b}\sum_{j\in b}\sum_{i = 0}^N d_i C_{ij}^{-1} (r_o^{\ell m}(b)+r_1^{\ell m}(b)(t_j-t_c)) h^{\ell m}_{o,j}\\
        &=\sum_b r_o^{\ell m}(b)\underbrace{\sum_{j\in b}\sum_{i = 0}^N d_i C_{ij}^{-1} h^{\ell m}_{o,j}}_{A_o^{\ell m}(b)}+\sum_b r_1^{\ell m}(b)\underbrace{\sum_{j\in b}\sum_{i = 0}^N d_i C_{ij}^{-1} h^{\ell m}_{o,j} (t_j-t_c)}_{A_1^{\ell m}(b)}\\
        &= \sum_b \left[r_o^{\ell m}(b)A_o^{\ell m}(b)+r_1^{\ell m}(b)A_1^{\ell m}(b)\right].
    \end{split}\label{eq:L_d_hlm}
\end{equation}
Similarly, we modify the expression for $L_1(s,s)$,
\begin{equation}
    \begin{split}
        L_1(s,s) &= L_1(F_+h_++F_\times h_\times,F_+h_++F_\times h_\times)\\
        &= F_+^2 L_1(h_+,h_+) + 2F_+F_\times L_1(h_+,h_\times) + F_\times^2 L_1(h_\times,h_\times)\\
        &= F_+^2 L_1(\mathfrak{Re}(\mathcal{H}),\mathfrak{Re}(\mathcal{H})) - 2F_+F_\times L_1(\mathfrak{Re}(\mathcal{H}),\mathfrak{Im}(\mathcal{H})) + F_\times^2 L_1(\mathfrak{Im}(\mathcal{H}),\mathfrak{Im}(\mathcal{H})).
    \end{split}\label{eq:L_s_s}
\end{equation}

Before moving forward, note that for any 2 complex numbers $z_1$ and $z_2$, we have
\begin{equation}
    \begin{split}
        \mathfrak{Re}(z_1)\mathfrak{Re}(z_2) &= \frac{\mathfrak{Re}(z_1\cdot z_2+\Bar{z}_1\cdot z_2)}{2},\\
        \mathfrak{Re}(z_1)\mathfrak{Im}(z_2) &= \frac{\mathfrak{Im}(z_1\cdot z_2+\Bar{z}_1\cdot z_2)}{2},\\
        \mathfrak{Im}(z_1)\mathfrak{Im}(z_2) &= \frac{\mathfrak{Re}(\Bar{z}_1\cdot z_2-z_1\cdot z_2)}{2}.
    \end{split}\label{eq:c_num_relations}
\end{equation}
We use these relations to simplify each term in \cref{eq:L_s_s}, starting with $F_+^2 L_1(\mathfrak{Re}(\mathcal{H}),\mathfrak{Re}(\mathcal{H}))$:
\begin{equation}
    \begin{split}
        F_+^2 L_1(\mathfrak{Re}(\mathcal{H}),\mathfrak{Re}(\mathcal{H})) &= F_+^2 \;\sum_{ij} \underbrace{\mathfrak{Re}(\mathcal{H}_i)}_{\mathfrak{Re}(z_1)} \underbrace{C_{ij}^{-1}\mathfrak{Re}(\mathcal{H}_j)}_{\mathfrak{Re}(z_2)}\\
        &= \frac{F_+^2}{2} \;\sum_{ij} \mathfrak{Re}\left[\mathcal{H}_i C_{ij}^{-1}\mathcal{H}_j+\mathcal{H}_i^* C_{ij}^{-1}\mathcal{H}_j\right]\\
        &= \frac{F_+^2}{2}\; \mathfrak{Re}\left[L_1(\mathcal{H},\mathcal{H}) + L_1(\mathcal{H}^*,\mathcal{H})\right].
    \end{split}\label{eq:Re_Re}
\end{equation}
Similarly,
\begin{equation}
    \begin{split}
        2F_+F_\times L_1(\mathfrak{Re}(\mathcal{H}),\mathfrak{Im}(\mathcal{H})) &= 2F_+F_\times \;\sum_{ij} \underbrace{\mathfrak{Re}(\mathcal{H}_i)}_{\mathfrak{Re}(z_1)} \underbrace{C_{ij}^{-1}\mathfrak{Im}(\mathcal{H}_j)}_{\mathfrak{Im}(z_2)}\\
        &= \frac{2F_+F_\times}{2} \;\sum_{ij} \mathfrak{Im}\left[\mathcal{H}_i C_{ij}^{-1}\mathcal{H}_j+\mathcal{H}_i^* C_{ij}^{-1}\mathcal{H}_j\right]\\
        &= F_+F_\times\; \mathfrak{Im}\left[L_1(\mathcal{H},\mathcal{H}) + L_1(\mathcal{H}^*,\mathcal{H})\right],
    \end{split}\label{eq:Re_Im}
\end{equation}
and 
\begin{equation}
    \begin{split}
        F_\times^2 L_1(\mathfrak{Im}(\mathcal{H}),\mathfrak{Im}(\mathcal{H})) &= F_\times^2 \;\sum_{ij} \underbrace{\mathfrak{Im}(\mathcal{H}_i)}_{\mathfrak{Im}(z_1)} \underbrace{C_{ij}^{-1}\mathfrak{Im}(\mathcal{H}_j)}_{\mathfrak{Im}(z_2)}\\
        &= \frac{F_\times^2}{2} \;\sum_{ij} \mathfrak{Re}\left[\mathcal{H}_i^* C_{ij}^{-1}\mathcal{H}_j-\mathcal{H}_i C_{ij}^{-1}\mathcal{H}_j\right]\\
        &= \frac{F_\times^2}{2}\; \mathfrak{Re}\left[L_1(\mathcal{H}^*,\mathcal{H}) - L_1(\mathcal{H},\mathcal{H})\right].
    \end{split}\label{eq:Im_Im}
\end{equation}
Substituting \cref{eq:Re_Re,eq:Re_Im,eq:Im_Im} into \cref{eq:L_s_s}, we obtain 
\begin{equation}
    \begin{split}
        L_1(s,s) = &\;\frac{F_+^2+F_\times^2}{2}\;\mathfrak{Re}( L_1(\mathcal{H}^*,\mathcal{H}))+\frac{F_+^2-F_\times^2}{2}\;\mathfrak{Re}(L_1(\mathcal{H},\mathcal{H}))\\
        &\; -F_+F_\times \mathfrak{Im}\left[L_1(\mathcal{H},\mathcal{H})+L_1(\mathcal{H}^*,\mathcal{H})\right].
    \end{split} \label{eq:final_eq:L_s_s}
\end{equation}
Each term in the above expression is computed using equation \cref{eq:pol_in_terms_of_modes} as
\begin{equation}
    \begin{split}
        L_1(\mathcal{H},\mathcal{H}) = \sum_{\ell_1,m_1>0}\sum_{\ell_2,m_2>0} & \big[\; _{-2}Y^{\ell_1,m_1} \;_{-2}Y^{\ell_2,m_2}L_1(h^{\ell_1m_1},h^{\ell_2m_2}) \\& + _{-2}Y^{\ell_1,m_1} \;_{-2}Y^{\ell_2,-m_2}(-1)^{\ell_2} L_1(h^{\ell_1m_1},h^{*\;\ell_2m_2}) \\ &+_{-2}Y^{\ell_1,-m_1} \;_{-2}Y^{\ell_2,m_2}(-1)^{\ell_1}L_1(h^{*\;\ell_1m_1},h^{\ell_2m_2})\\ &+_{-2}Y^{\ell_1,-m_1} \;_{-2}Y^{\ell_2,-m_2}(-1)^{\ell_1+\ell_2}L_1(h^{\ell_1m_1},h^{\ell_2m_2})^*\big],
    \end{split}\label{eq:L_H_H}
\end{equation}
and
\begin{equation}
    \begin{split}
        L_1(\mathcal{H}^*,\mathcal{H}) = \sum_{\ell_1,m_1>0}\sum_{\ell_2,m_2>0} & \big[\; _{-2}Y^{*\;\ell_1,m_1} \;_{-2}Y^{\ell_2,m_2}L_1(h^{*\;\ell_1m_1},h^{\ell_2m_2}) \\ &+ _{-2}Y^{*\;\ell_1,m_1} \;_{-2}Y^{\ell_2,-m_2}(-1)^{\ell_2} L_1(h^{\ell_1m_1},h^{\ell_2m_2})^* \\ &+_{-2}Y^{\ell_1,-m_1} \;_{-2}Y^{\ell_2,m_2}(-1)^{\ell_1}L_1(h^{\ell_1m_1},h^{\ell_2m_2})\\&+_{-2}Y^{*\;\ell_1,-m_1} \;_{-2}Y^{\ell_2,-m_2}(-1)^{\ell_1+\ell_2}L_1(h^{\ell_1m_1},h^{*\;\ell_2m_2})\big].
    \end{split}\label{eq:L_H*_H}
\end{equation}
The terms $L_1(h^{\ell m},h^{\ell m})$ and $L_1(h^{\ell m},h^{*\;\ell m})$ are further written in terms of fiducial modes $h^{\ell m}_o$ and ratio of waveform modes $r^{\ell m}$ as, 
\begin{equation}
    \begin{split}
        \Rightarrow L_1(h^{\ell_1m_1},h^{\ell_2m_2}) =  \sum_{b_1}\sum_{b_2}& r_o^{\ell_1m_1}(b_1) r_o^{\ell_2m_2}(b_2)\underbrace{ \sum_{i\in b_1}\sum_{j\in b_2} h^{\ell_1m_1}_{o,i}\; C_{ij}^{-1}\; h^{\ell_2m_2}_{o,j}}_{B_o^{\ell_1m_1,\ell_2m_2}(b_1,b_2)}\\
         +\sum_{b_1}\sum_{b_2}& r_o^{\ell_1m_1}(b_1)r_1^{\ell_2m_2}(b_2) \underbrace{\sum_{i\in b_1}\sum_{j\in b_2}h^{\ell_1m_1}_{o,i}\; C_{ij}^{-1}\; h^{\ell_2m_2}_{o,j}(t_j-t_c(b_2))}_{B_1^{\ell_1m_1,\ell_2m_2}(b_1,b_2)}\\
         +\sum_{b_1}\sum_{b_2}& r_o^{\ell_2m_2}(b_1) r_1^{\ell_1m_1}(b_2) \underbrace{\sum_{i\in b_1}\sum_{j\in b_2}h^{\ell_2m_2}_{o,i} \; C_{ij}^{-1}\;h^{\ell_1m_1}_{o,j} (t_j-t_c(b_1))}_{B_2^{\ell_1m_1,\ell_2m_2}(b_1,b_2)},
    \end{split}\label{eq:L_1_hl1m1_hl2m2}
\end{equation}
\begin{equation}
    \begin{split}
        L_1(h^{*\;\ell_1m_1},h^{\ell_2m_2})
         = \sum_{b_1}\sum_{b_2} &r_o^{*\;\ell_1m_1}(b_1) r_o^{\ell_2m_2}(b_2)\underbrace{ \sum_{i\in b_1}\sum_{j\in b_2} h^{*\;\ell_1m_1}_{o,i}\; C_{ij}^{-1}\; h^{\ell_2m_2}_{o,j}}_{B_3^{\ell_1m_1,\ell_2m_2}(b_1,b_2)}\\
         +\sum_{b_1}\sum_{b_2} &r_o^{*\;\ell_1m_1}(b_1)r_1^{\ell_2m_2}(b_2) \underbrace{\sum_{i\in b_1}\sum_{j\in b_2}h^{*\;\ell_1m_1}_{o,i}\; C_{ij}^{-1}\; h^{\ell_2m_2}_{o,j}(t_j-t_c(b_2))}_{B_4^{\ell_1m_1,\ell_2m_2}(b_1,b_2)}\\
         +\sum_{b_1}\sum_{b_2}& r_o^{\ell_2m_2}(b_1) r_1^{*\;\ell_1m_1}(b_2) \underbrace{\sum_{i\in b_1}\sum_{j\in b_2}h^{\ell_2m_2}_{o,i}\; C_{ij}^{-1}\;h^{*\;\ell_1m_1}_{o,j} (t_j-t_c(b_1))}_{B_5^{\ell_1m_1,\ell_2m_2}(b_1,b_2)},
    \end{split}\label{eq:L_1_h*l1m1_hl2m2}
\end{equation}
and 
\begin{equation}
    \begin{split}
        L_1(h^{\ell_1m_1},h^{*\;\ell_2m_2})
         = \sum_{b_1}\sum_{b_2} &r_o^{\ell_1m_1}(b_1) r_o^{*\;\ell_2m_2}(b_2)\underbrace{ \sum_{i\in b_1}\sum_{j\in b_2} h^{\ell_1m_1}_{o,i}\; C_{ij}^{-1}\; h^{*\;\ell_2m_2}_{o,j}}_{B_6^{\ell_1m_1,\ell_2m_2}(b_1,b_2)}\\
         +\sum_{b_1}\sum_{b_2}& r_o^{\ell_1m_1}(b_1)r_1^{*\;\ell_2m_2}(b_2) \underbrace{\sum_{i\in b_1}\sum_{j\in b_2}h^{\ell_1m_1}_{o,i}\; C_{ij}^{-1}\; h^{*\;\ell_2m_2}_{o,j}(t_j-t_c(b_2))}_{B_7^{\ell_1m_1,\ell_2m_2}(b_1,b_2)}\\
         +\sum_{b_1}\sum_{b_2}& r_o^{*\;\ell_2m_2}(b_1) r_1^{\ell_1m_1}(b_2) \underbrace{\sum_{i\in b_1}\sum_{j\in b_2}h^{*\;\ell_2m_2}_{o,i}\; C_{ij}^{-1}\;h^{\ell_1m_1}_{o,j} (t_j-t_c(b_1))}_{B_8^{\ell_1m_1,\ell_2m_2}(b_1,b_2)}.
    \end{split}\label{eq:L_1_hl1m1_h*l2m2}
\end{equation}

The quantities $A_n^{\ell m}(b)$ and $B_n^{\ell_1m_1, \ell_2 m_2}(b_1,b_2)$ introduced in \cref{eq:L_d_hlm,eq:L_1_hl1m1_hl2m2,eq:L_1_h*l1m1_hl2m2,eq:L_1_hl1m1_h*l2m2}, respectively, are collectively known as \textit{summary data}. 

\section{Efficient Likelihood computation with symmetric Toeplitz covariance matrices}\label{apx:Efficient_likelihood_GS_method}

The evaluation of time-domain Gaussian likelihood, defined in \cref{eq:TD_likelihood_single_detector}, requires inverse of the noise covariance matrix. A direct inversion of an $N\times N$ matrix is computationally expensive and scales as $\mathcal{O}(N^3)$. Further, storing such a matrix requires $\mathcal{O}(N^2)$ memory, which can be orders of terabytes for a 128-second-long data segment. 

However, for stationary random Gaussian noise, the covariance matrix simplifies to a symmetric Toeplitz form (see \cref{s2:bayesian_pe}). We exploit this structure to reduce the memory requirements and computational cost of likelihood evaluation. We employ two distinct methods to handle these matrix operations: the Levinson-Durbin recursion and the Gohberg-Semencul inversion method~\cite {Gohberg2010}.

\begin{enumerate}
    \item \textbf{Levinson-Durbin recursion method:} We utilise the \texttt{solve\_toeplitz} function from \texttt{scipy}~\cite{virtanen2020scipy} to compute $C^{-1} h$. This function solves a Toeplitz system of form $Tx = y$ using Levinson-Durbin recursion and requires only the first row and first column of the Toeplitz matrix, which reduces the memory cost to $\mathcal{O}(N)$. However, the complexity of this method scales as $\mathcal{O}(N^2)$, which makes it inefficient for longer data segments. 
    
    \item \textbf{Gohberg-Semencul inversion method:} To overcome the complexity of \texttt{solve\_toeplitz} function, we employ Gohberg-Semencul inversion method. This formula expresses the inverse of a Toeplitz matrix as the difference of products of triangular Toeplitz matrices. For a Toeplitz matrix $A$, the inverse is given by 
    \begin{equation}
        A^{-1} = \frac{1}{x_0}\left\{\begin{bmatrix}
            x_0 & 0 & \cdots & 0 \\ 
            x_1 & x_0 & \ddots & \vdots \\ 
            \vdots & \vdots & \ddots & 0\\
            x_n & x_{n-1} & \cdots & x_0
        \end{bmatrix} \begin{bmatrix}
            y_0 & y_{-1} & \cdots & y_{-n} \\ 
            0 & y_0 & \cdots & y_{1-n} \\ 
            \vdots & \ddots & \ddots & \vdots\\
            0  & \cdots  & 0& y_0 
        \end{bmatrix} 
         - \begin{bmatrix}
            0 & 0 & \cdots & 0 \\ 
            y_{-n} & 0 & \ddots & 0 \\ 
            \vdots & \vdots & \ddots & 0\\
            y_{-1} & y_{-2} & \cdots & 0
        \end{bmatrix} \begin{bmatrix}
            0 & x_{n} & \cdots & x_1 \\ 
            0 & 0 & \cdots & x_2 \\ 
            \vdots & \ddots & \ddots & \vdots\\
            0  & \cdots  & 0& 0 
        \end{bmatrix}\right\},
        \label{eq:GS_inversion}
    \end{equation}
    where $\{x_i, y_i\}$ are solutions to the linear equations
    \begin{equation}
    \begin{split}
        A \begin{bmatrix}
            x_0 \\ x_1 \\ \vdots \\ x_n
        \end{bmatrix}
          = \begin{bmatrix}
             1 \\ 0 \\ \vdots \\ 0
         \end{bmatrix}, \;\;\;\;\;
         \begin{bmatrix}
             y_{-n} & \cdots & y_{-1} & y_0
         \end{bmatrix}A = \begin{bmatrix}
             0 & \cdots & 0 & 1
         \end{bmatrix}.
    \end{split}
    \end{equation}
    For a symmetric Toeplitz matrix, we only need to compute $x$ by solving above linear equations and evaluate vector $y$ using $y_k = x_{n-k}$. 

    In our implementation, instead of constructing a full inverse covariance matrix, we compute and store only the vectors $x$ and $y$ using \texttt{solve\_toeplitz}, which reduces the memory cost to $\mathcal{O}(N)$. To evaluate the term $C^{-1}h$, we compute the product of each term in \cref{eq:GS_inversion} with vector $h$ separately. We use the \texttt{matmul\_toeplitz} function to compute these products using only the vectors $x$, $y$ and $h$. This function performs matrix operations using FFT, which reduces the computational complexity to $\mathcal{O}(N\log N)$. In \cref{fig:Likelihood_comparison_two_methods}, we compare the computational efficiency of both methods for a single likelihood calculation.

    \begin{figure}[htb]
    \centering
    \includegraphics[width=0.6\linewidth]{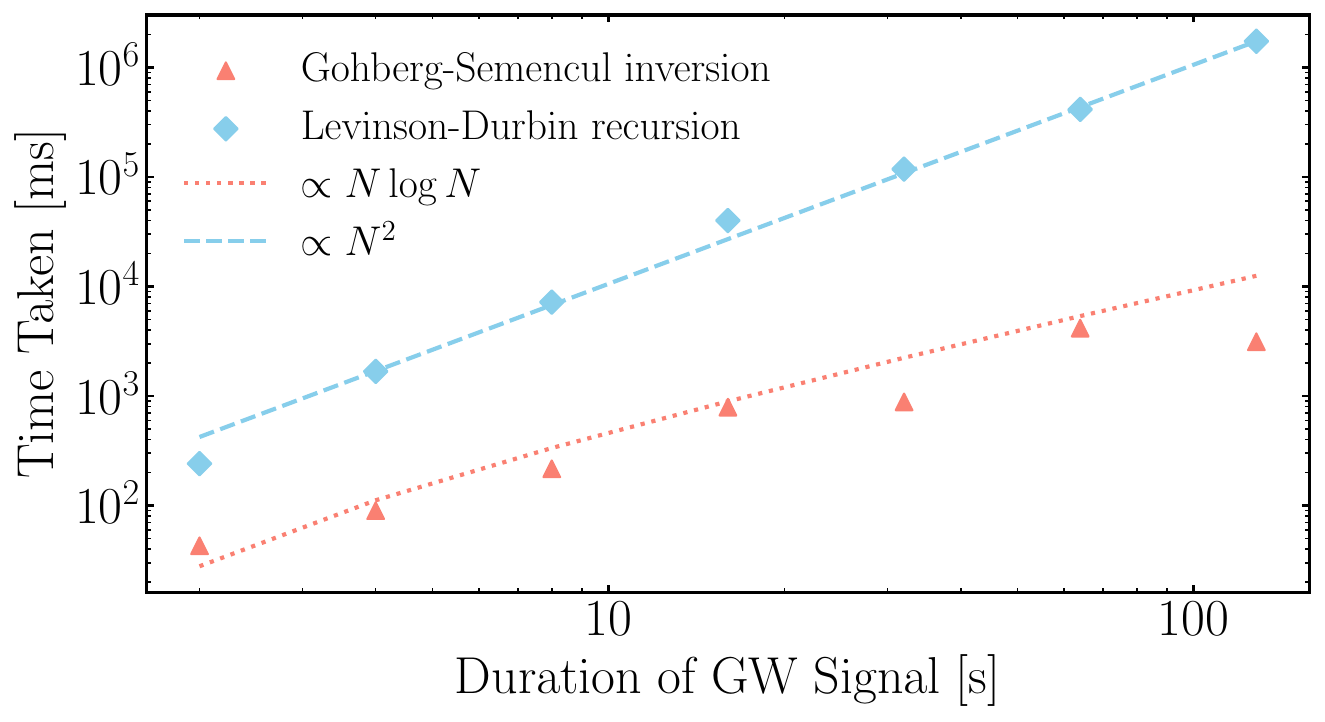}
    \caption{Comparison of time taken for a single likelihood evaluation using Levinson-Durbin recursion and Gohberg-Semencul inversion methods for GW signals of varying duration. The time taken by Levinson-Durbin recursion method scales as $\mathcal{O}(N^2)$, while Gohberg-Semencul inversion method scales as $\mathcal{O}(N\log N)$. For longer signals, the Gohberg-Semencul method is computationally efficient.}
    \label{fig:Likelihood_comparison_two_methods}
    \end{figure}

\end{enumerate}

\section{Effect of insufficient bins}
\begin{figure}[htb]
    \centering
    \includegraphics[width=1.0\linewidth]{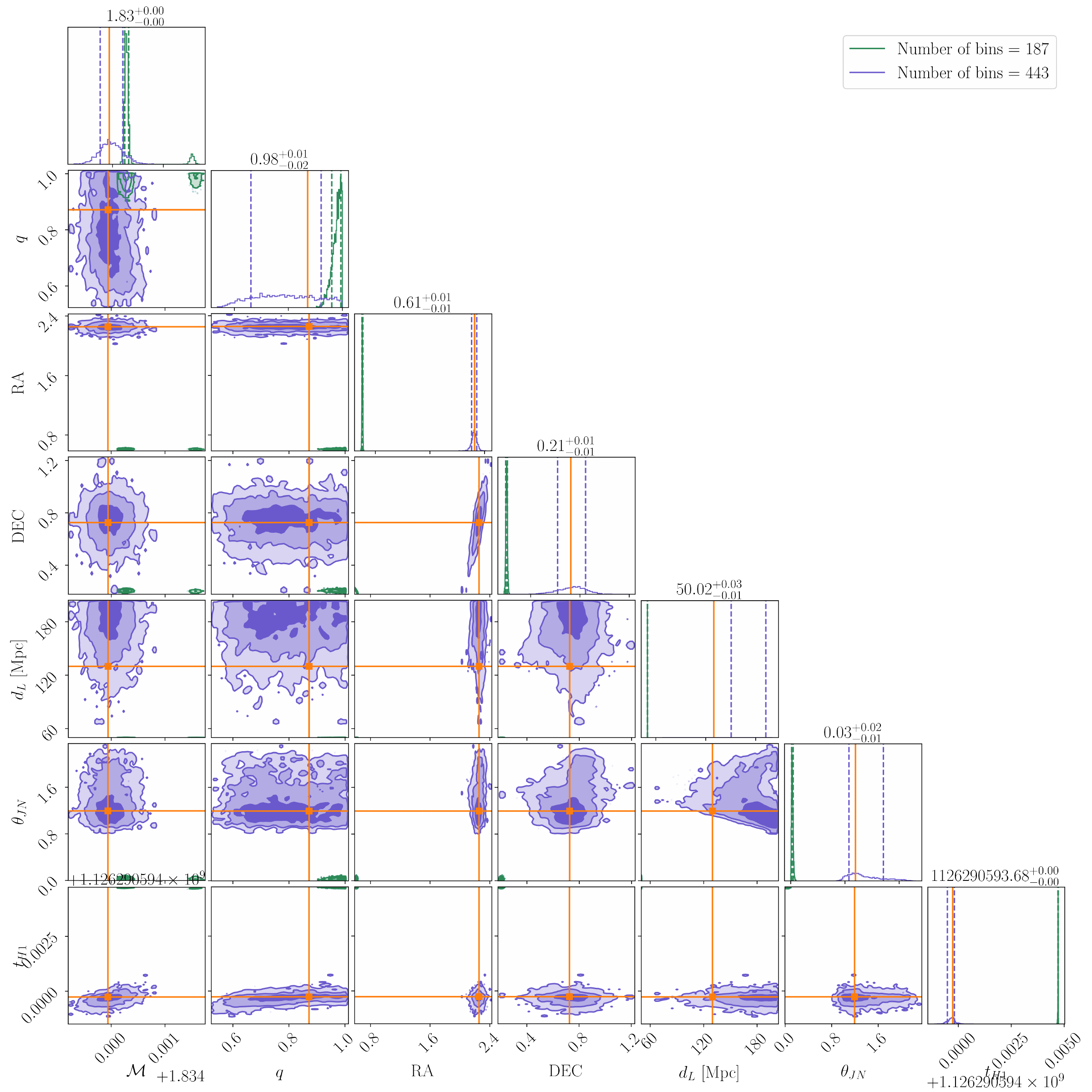}
    \caption{Comparison between the posterior probability distributions for a 128-second long signal, estimated using heterodyning with 443 bins (purple) and 187 bins (green). The analysis with 443 bins accurately recovers the posteriors, but analysis with 187 causes linear approximation of waveform modes to break down, resulting in biased posteriors.}
    \label{fig:Comparison_between_diff_number_of_bins}
    \end{figure}

The accuracy of heterodyning method relies on the assumption that the ratio of waveform modes can be approximated as a linear function within a small time bin. The total number and width of bins are chosen using a binning criteria derived in \cref{s2:Binning_criteria}. If the bins are too wide, the linear approximation of ratio of modes breaks down, leading to a false inference. In \cref{fig:Comparison_between_diff_number_of_bins}, we show the consequence of choosing too few bins when using heterodyning for parameters estimation. 
\end{document}